\documentclass{ifacconf}

\usepackage{graphicx}      
\usepackage{natbib}        
\usepackage{times} 
\usepackage{amsmath} 
\usepackage{amssymb}  
\usepackage{pifont}
\usepackage{algorithm}
\usepackage{graphicx}
\usepackage{subcaption}
\usepackage{xcolor}
\usepackage{algpseudocode}
\usepackage{url}
\usepackage{placeins}
\newcommand{\zono}[1]{\left\langle#1\right\rangle}
\newcommand{\cZono}[1]{\left\langle#1\right\rangle}

\newcommand{\confSet}{\mathcal{H}_t}

\newcommand{\interval}[2]{\left[ #1,#2 \right]}
\newcommand{\Rspace}{\mathbb{R}}
\newcommand{\idmatrix}{\mathbf{I}}
\newcommand{\zeros}{\mathbf{0}}
\newcommand{\ones}{\mathbf{1}}
\newcommand{\unitvector}[2]{[\zeros_{1\times(i-1)}\; 1\; \zeros_{1\times(n-i)}]}

\newcommand{\vol}[1]{\textbf{vol}\left(#1\right)}

\usepackage{etoolbox}

\begin{document}
\begin{frontmatter}

\title{Shared Situational Awareness using\\Hybrid Zonotopes with Confidence Metric\thanksref{ack}} 

\author[KTH,Traton]{Vandana Narri\thanksref{info1}}
\author[PSU]{Jonah J. Glunt\thanksref{info1}}
\author[PSU]{Joshua A. Robbins}
\author[KTH]{Jonas Mårtensson}
\author[PSU]{Herschel C. Pangborn}
\author[KTH]{Karl H. Johansson}%
\address[KTH]{KTH Royal Institute of Technology. {\tt\small\{narri, jonas1, kallej\}@kth.se.}}%
\address[Traton]{Scania AB CVTraton Group, Sweden. {\tt\small vandana.narri@scania.com.}}
\address[PSU]{The Pennsylvania State~University. {\tt\small\{jglunt, jrobbins, hcpangborn\}@psu.edu.}}%
\thanks[ack]{This work was partially supported by the Wallenberg Artificial Intelligence, Autonomous Systems, and Software Program (WASP) funded by the Knut and Alice Wallenberg Foundation. It was also partially supported by the Swedish Research Council Distinguished Professor Grant 2017-01078 and the Knut and Alice Wallenberg Foundation Wallenberg Scholar Grant. The work from Jonah Glunt was supported by the U.S. Department of Defense through the National Defense Science \& Engineering Graduate (NDSEG) Fellowship Program.}%
\thanks[info1]{The first two authors contributed equally to this research.}

\begin{abstract}
Situational awareness for connected and automated vehicles describes the ability to perceive and predict the behavior of other road-users in the near surroundings. However, pedestrians can become occluded by vehicles or infrastructure, creating significant safety risks due to limited visibility. Vehicle-to-everything communication enables the sharing of perception data between connected road-users, allowing for a more comprehensive awareness. The main challenge is how to fuse perception data when measurements are inconsistent with the true locations of pedestrians. Inconsistent measurements can occur due to sensor noise, false positives, or unmodeled disturbances. This paper employs set-based estimation with constrained zonotopes to compute a confidence metric for the measurement set from each sensor. Estimated sets and their confidences are then fused using hybrid zonotopes. This method can account for inconsistent measurements, enabling reliable and robust fusion of the sensor data. The effectiveness of the proposed method is demonstrated in both simulation and real experiments.
\end{abstract}

\begin{keyword}
Set-Based Estimation, Connected and Automated Vehicles, Collaborative Perception, Pedestrian Safety.
\end{keyword}

\end{frontmatter}
\section{Introduction}
One of the main challenges in the development of connected and automated vehicles~(CAVs) for urban scenarios is limited visibility due to occlusions in their field-of-view (FOV). Most occlusions are caused by moving vehicles, parked vehicles, and infrastructure. Among road-users, pedestrians are the most vulnerable in terms of safety. In~\cite{WHO2023global}, a global status report on road safety situation in 194 countries highlights that 56\% of deaths are among unprotected road-users such as pedestrians, cyclists, and motorcyclists. As proposed by~\cite{narri2021set, narri2023shared, narri2025situational}, vehicle-to-everything~(V2X) communication can be leveraged to enable shared situational awareness through the sharing of perception data among V2X units, defined as any CAV or road-side unit (RSU) capable of communicating within this framework. However, due to sensor noise or unmodeled disturbances, measurements from V2X units can be inconsistent with each other. This introduces a new challenge in fusing data across V2X units.

\begin{figure}
    \centering
    \includegraphics[trim ={1.3cm 2.8cm 1.4cm 3.1cm},clip,width=0.9\linewidth]{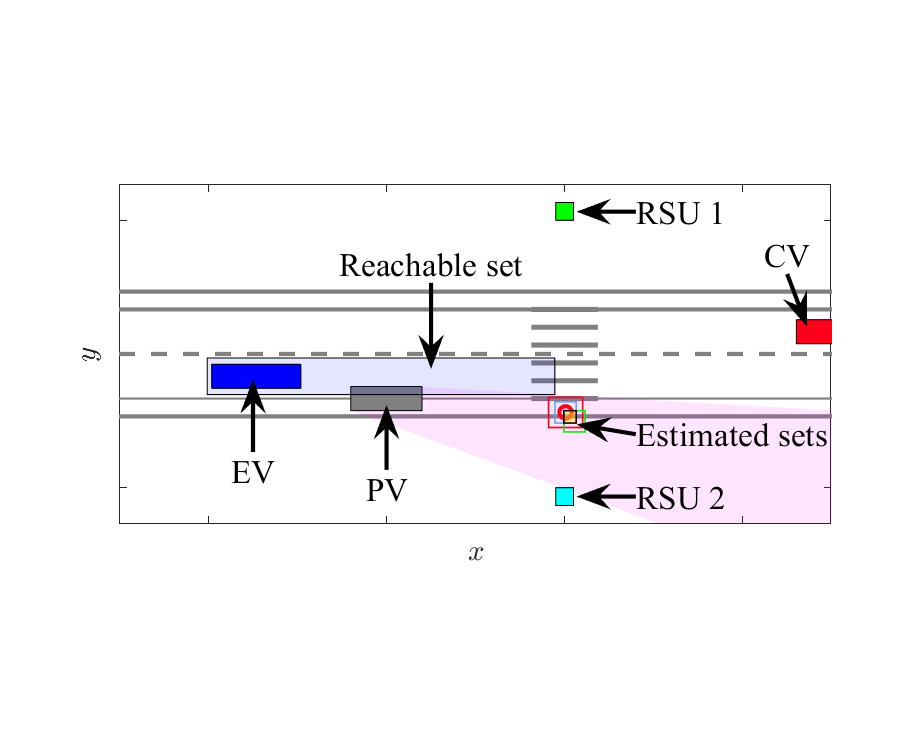}
    \caption{Snapshot of the simulated occluded pedestrian-crossing scenario with inconsistent measurement sets.}
    \label{fig:MATLAB_sim_overview}
\end{figure}

This work proposes a set-based estimation and fusion method that encodes measurement inconsistencies in a confidence metric to increase pedestrian safety. Consider the urban scenario illustrated in Fig.~\ref{fig:MATLAB_sim_overview}. The ego vehicle~(EV) is moving from left to right, while a parked vehicle~(PV) to its right creates an occluded region in the EV's FOV, highlighted by the magenta shaded region. The set of states reachable by the EV within the next $2$s, referred to as the reachable set, is highlighted in light blue on the road. A pedestrian, indicated by the red circle, is located in this occluded region and about to cross the road. A connected vehicle~(CV) moving from right to left is able perceive the pedestrian. Additionally, there are two road-side units~(RSUs) that can also see the pedestrian. The CV, RSU~1, and RSU~2, each produce measurement sets and calculate estimated sets intended to bound the pedestrian's location, given by the red, green, and blue boxes, respectively.
The intersection of these three sets is highlighted in yellow. When there is noise or bias in the measured sets that violates the assumed bounds used to make the estimated sets, this intersection can become very small or even empty, which can cause existing set-based estimation techniques to yield an empty fused set, effectively losing all the measurement information.
The methods proposed in this paper allow the EV to construct a fused set from the measurements received from each sensor without losing any relevant information in the case of inconsistencies.

\subsection{Literature Review}

Many recent contributions use V2X communication to improve EV's situational awareness.~\cite{ngo2023cooperative} and~\cite{yee2018collaborative} emphasize that one way is to create collaborative perception by sharing object detections and using neural networks to fuse the data.~\cite{CP_methods_datasets_challenges} provide a detailed survey of existing collaborative perception methods and their challenges of  when incorporating in traffic scenarios. They highlight real-world issues, such as communication imperfections and delays, localization errors, model discrepancies, and security issues, which can each deteriorate the system's safety and precision. Sharing information between CAVs is expected to increase the efficiency, safety, and capacity of transportation systems. However, the increasing usage of shared sensor information can also increase the vulnerability of CAVs to sensor faults and adversarial attacks.
\cite{Sensor_Failures_in_Autonomous_Vehicles} surveys sensor and fusion technologies and highlights serval weaknesses of these methods that can affect road-users safety in urban scenarios. A sensor reading can be degraded for a variety of reasons, such as signal outages, poor weather conditions, other environmental circumstances, or poor sensor quality.  \cite{sensor_noise_factors} present a thorough analysis of different types of sensor noise and their effects on sensor output.

We propose to use set-based methods to improve the robustness of V2X communication to measurement noise and uncertainty. Set-based estimators have also been used in fault detection~\citep{conf:set_fault1}, underwater robotics~\citep{jaulin2009nonlinear}, ground vehicles~\citep{franze2015obstacle}, multi-agent systems~\citep{garcia2020guaranteed}, and localization~\citep{conf:setloc}.
\cite{ScottJosephKConstrainedZonotopes} introduced constrained zonotopes as a computationally efficient set representation for convex polytopes. 
Constrained zonotopes have found great use in set-valued state estimation~\citep{ScottJosephKConstrainedZonotopes,  raimondo2016closed, rego2020guaranteed} due to their efficient identities for common set operations. 
They have been applied to other problems as well, such as robust controllable and robust positively invariant set computations~\citep{vinod2025projection, raghuraman2022set}, hierarchical control~\citep{koeln2020vertical}, and tube-based model predictive control~\citep{andrade2024tube}. Hybrid zonotopes~\citep{hybrid2023} extend constrained zonotopes by incorporating integer variables. This allows for representation of nonconvex and disjoint sets, enabling nonlinear state estimation~\citep{siefert2023_SVSE, thompson2025_MHE}.
In this study, estimated sets are represented by constrained zonotopes and fused sets are represented by hybrid zonotopes, as the proposed methods involve unions. 
This work uses the zonoLAB toolbox~\citep{koeln2024zonolab} for storing and representing zonotopic sets in MATLAB for simulations and the ZonoOpt toolbox~\citep{robbins2025sparsity_zonoopt} for operations in C++ for experiments. 

\subsection{Contributions}
This work extends the shared situational awareness framework proposed in~\cite{narri2025situational} by addressing scenarios with inconsistent measurement sets. 
The specific contributions of this work are: 
\begin{enumerate}
    \item A confidence metric derived from previous and current estimated sets, which can be computed alongside a set-based estimation of the pedestrian location. 
    \item A set-based fusion method that combines estimated sets for V2X units without losing any relevant information.
    \item An analysis of effectiveness of this method in simulations.
    \item An analysis and discussion of testing of this proposed method using real-time data collected from Scania's connected and automated prototype vehicles.
\end{enumerate}

\subsection{Outline}
\label{subsec:outline}
The remainder of this paper is organized as follows. 
Section~\ref{sec:preliminaries} gives mathematical preliminaries for constrained and hybrid zonotopes. The proposed framework and algorithm are described in Section~\ref{sec:method}. Section~\ref{sec:simulation_setup_results} presents simulation results, followed by real-world experiments in Section~\ref{sec:real_setup_results}. Section~\ref{sec:conclu} concludes the paper.
\section{Preliminaries}
\label{sec:preliminaries}
This section defines the mathematical notation and set representations used in this paper.
The interval $\mathcal{I} =\{x\in\Rspace^\gamma \mid a \leq x \leq b\}$ is denoted by $\interval{a}{b}$. The $\gamma$-dimensional identity matrix is given by $\idmatrix_\gamma$, and a matrix of all zeros (respectively ones) is given by $\zeros$ (respectively $\ones$) with dimensions provided in subscript when not clear from context. 

A constrained zonotope $\mathcal{X} \subset \mathbb{R}^\gamma$~\citep{ScottJosephKConstrainedZonotopes} is a bounded convex polytope written in the form
\begin{equation}
     \mathcal{X} = \Big\{ x \in \mathbb{R}^\gamma \Big| x = h + G\beta , A \beta = b, \,\beta\in\interval{-1}{1}^e \Big\},
    \label{eq:cZonoDef}
\end{equation}%
where $h \in \mathbb{R}^\gamma$ is the center and $G \in \mathbb{R}^{\gamma \times e}$ is the generator matrix of the constrained zonotope. The matrix $A \in \mathbb{R}^{\gamma_c \times e}$ and the vector $b \in \mathbb{R}^{\gamma_c}$ define the constraints. 
The shorthand notation of a constrained zonotope is $\mathcal{X}=\cZono{h,G,A,b}$.
This work utilizes constrained zonotopes to take advantage of their efficient, closed-form expressions for common set operations. Given two constrained zonotopes $\mathcal{X}_1 = \cZono{h_1,G_1,A_1,b_1}$ and $\mathcal{X}_2 = \cZono{h_2,G_2,A_2,b_2}$, then
\begin{subequations} \label{eq:set-ops-conZono}
\begin{align}
&R \mathcal{X}_1 + s = \left\langle R h_1 + s, RG_1, A_1, b_1 \right\rangle, \label{eq:set-ops-lin-map} \\
&\mathcal{X}_1 \times \mathcal{X}_2 = \left\langle \begin{bmatrix}
    h_1 \\ h_2
\end{bmatrix}, \begin{bmatrix}
    G_1 & \zeros \\
    \zeros & G_2
\end{bmatrix},  \begin{bmatrix}
    A_1 & \zeros \\
    \zeros & A_2
\end{bmatrix}, \begin{bmatrix}
    b_1 \\ b_2
\end{bmatrix} \right\rangle, \label{eq:set-ops-cart-prod} \\
&\mathcal{X}_1 \oplus \mathcal{X}_2 = \left\langle \!h_1 + h_2, \begin{bmatrix}
    G_1 & G_2
\end{bmatrix}, \begin{bmatrix}
    A_1 & \zeros \\
    \zeros & A_2
\end{bmatrix}, \begin{bmatrix}
    b_1 \\ b_2
\end{bmatrix} \!\right\rangle , \label{eq:set-ops-mink-sum} \\
&\mathcal{X}_1 \cap_R \mathcal{X}_2 = \Bigg\langle h_1, \begin{bmatrix}
    G_1 & \zeros
\end{bmatrix}, \begin{bmatrix}
    A_1 & \zeros \\
    \zeros & A_2 \\
    R G_1 & -G_2
\end{bmatrix}, \begin{bmatrix}
    b_1 \\ b_2 \\ h_2 - R h_1
\end{bmatrix} \Bigg\rangle, \label{eq:set-ops-intersection}
\end{align}
\end{subequations}%
where \eqref{eq:set-ops-lin-map} is the affine map, \eqref{eq:set-ops-cart-prod} is the Cartesian product, \eqref{eq:set-ops-mink-sum} is the Minkowski sum, and \eqref{eq:set-ops-intersection} is the generalized intersection. Note that points can be represented as constrained zonotopes with empty generator and constraint matrices and used in these set operation identities. A zonotope is a constrained zonotope without constraints, and is written as $\mathcal{X} = \zono{h, G}$.

A hybrid zonotope $\mathcal{H} \subset \mathbb{R}^\gamma$~\citep{hybrid2023} is an extension of constrained zonotopes that includes binary generators, and is written as
\begin{align}
    \mathcal{H} =  \Big\{ x \in \mathbb{R}^\gamma \Big| & x = h + \begin{bmatrix}
        G^c & G^b
    \end{bmatrix} \begin{bmatrix}
        \beta^c \\ \beta^b
    \end{bmatrix},\nonumber \begin{bmatrix}
        A^c & A^b
    \end{bmatrix} \begin{bmatrix}
        \beta^c \\ \beta^b
    \end{bmatrix} = b,\nonumber\\&
    \begin{bmatrix}
        \beta^c \\ \beta^b
    \end{bmatrix} \in \interval{-1}{1}^e \times \{ -1, 1 \}^p \Big\},
    \label{eq:hybird_set}
\end{align}%
where $h \in \mathbb{R}^{\gamma}$, $G^c \in \mathbb{R}^{\gamma \times e}$, $G^b \in \mathbb{R}^{\gamma \times p}$, $A^c \in \mathbb{R}^{\gamma_c \times e}$, $A^b \in \mathbb{R}^{\gamma_c \times p}$, and $b \in \mathbb{R}^{\gamma_c}$. The generator and constraint
matrices have been partitioned into continuous and
binary parts denoted by superscripts $c$ and $b$, respectively, to
form the mixed-integer set representation. The superscript $b$
should not be confused with the vector $b$ used for the equality constraints. The shorthand
notation of a hybrid zonotope is $\mathcal{H} =\cZono{h,G^c,G^b, A^c, A^b, b}$. Hybrid zonotopes have similar identities for common set operations as those for constrained zonotopes in~\eqref{eq:set-ops-conZono}. These can be found in~\cite{hybrid2023}, while identities for other operations such as unions and complements can be found in~\cite{bird2021unions}. This paper utilizes hybrid zonotopes for their ability to efficiently represent unions of zonotopic sets, via the identity in~\cite{robbins2026_hybrid}.

For a set $\mathcal{X}\subset\Rspace^\gamma$, the $\gamma$-dimensional volume is denoted by $\vol{\mathcal{X}}$. To compute the volume of a constrained zonotope, first the vertices are enumerated using an iterative support-function based search. Vertex enumeration of a constrained zonotope has exponential time complexity in the worst case, just as converting from halfspace to vertex representation of polytopes~\citep{polyhedra_vertex_enumeration_Assad2022}. Then the volume can be computed from the vertices, using the \emph{surveyor's area formula} if in two dimensions~\citep{Braden01091986}, or more generally by dividing the set into a disjunction of simplices and summing their volumes. 

\begin{figure*}
    \centering
    \includegraphics[width=0.9\linewidth]{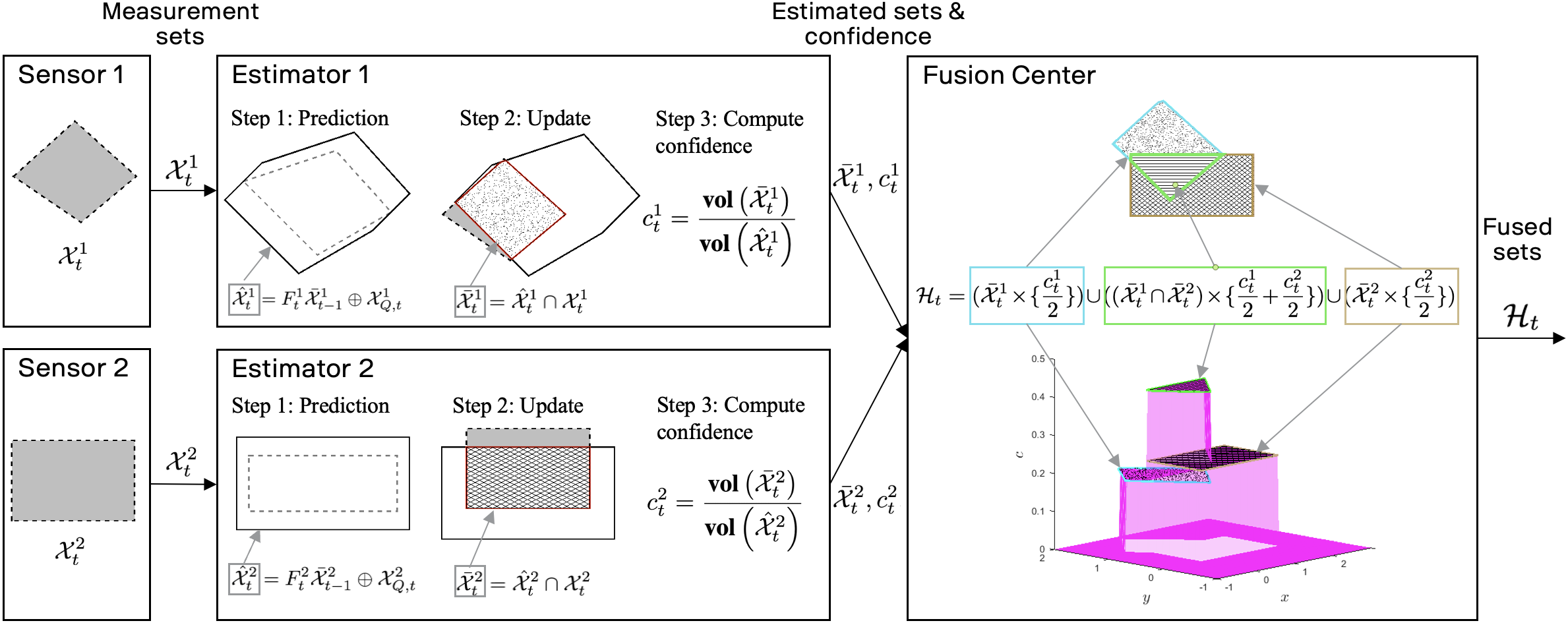}
    \caption{Architecture of the proposed framework when applied with two sensors. The output is the fused set~$\confSet$ for the detected pedestrian, computed using the two measurement sets,~$\mathcal{X}^1_t$ from sensor~1 and~$\mathcal{X}^2_t$ from sensor~2.}
    \label{fig:illustration_of_proposed_framework}
\end{figure*}
\section{Methodology}
\label{sec:method}

This section presents the architecture and algorithms of the proposed framework for shared situational awareness. Here, it is assumed that there is one pedestrian in the occluded region and this pedestrian is not visible to the EV but is detected by the V2X units and the CV, as presented in Fig.~\ref{fig:MATLAB_sim_overview}. Each V2X unit in the scenario is assumed to provide a measurement set. A measurement set at time $t$ is denoted by $\mathcal{X}_t \subset \mathbb{R}^\gamma$ for some $\gamma\geq1$. Sets are used to account for uncertainty in the velocity and heading of the pedestrian, along with bounded sensor noise. This noise is assumed to be random, unknown, and sensor-dependent. 
A key challenge is computing a robust fused set that combines the information from each V2X unit without becoming empty if the measurement sets are inconsistent.

\subsection{Architecture}

The proposed architecture consists of three types of modules: sensors, estimators, and a fusion center, as shown in Fig.~\ref{fig:illustration_of_proposed_framework} for a case with two sensors corresponding to two V2X units. Sensors 1 and 2 produce measurement sets~$\mathcal{X}^1_t$ and~$\mathcal{X}^2_t$, respectively, of the pedestrian in their FOV at discrete time step $t$. The measurement sets are then communicated to their respective estimators. Each estimator computes an estimated set $\bar{\mathcal{X}}^{i}_t$ and confidence metric~$c^i_t$ for the detected pedestrian using the previously estimated set at time $t-1$ and the measurement set $\mathcal{X}^i_t$. Here, superscript~$i = 1,2$ denotes the sensor index. The estimated sets and their confidence metrics are then communicated to the fusion center and used to construct a fused set~$\confSet$. This fused set can be passed to the motion planner of the EV. The proposed framework generalizes naturally from two to $n$ sensors.

\subsection{Estimation}

To obtain an estimated set for each sensor, we follow a standard set-based estimation procedure of performing a dynamic update of the previous estimated set and then intersecting with the current measurement set, as in~\citep{ScottJosephKConstrainedZonotopes}. A discrete-time linear system is considered to model the motion of the pedestrian as
\begin{equation}
    x_{t} = F_t x_{t-1} + q_{t},
    \label{eq:system_model}
\end{equation}
where $x_{t} \in \mathbb{R}^\gamma$ is the location of the pedestrian, $F_t \in \mathbb{R}^{\gamma \times \gamma}$ the transition matrix, and $q_t \in \mathbb{R}^\gamma$ the uncertainty of the pedestrian velocity and heading. By assuming a constant maximum velocity, the possible displacements in $x$ and $y$ can be computed, which represents the uncertainty in both the heading and velocity, and enables one-step prediction without imposing constraints on the pedestrian’s heading. The unknown initial position of the pedestrian is denoted by $x_0$ and assumed to be bounded by an initial estimated set $\bar{\mathcal{X}}_0 \subset \mathbb{R}^\gamma$. The uncertainties in the motion model are represented by a  zonotope: $q_t \in \mathcal{X}_{Q,t} =\zono{\zeros,Q_t}$. 
At time step $t$, given the $i^{th}$ sensor's previous estimate $\bar{\mathcal{X}}_{t-1}^i$ and the bounded uncertainty~$\mathcal{X}_{Q,t}$, we define the \emph{predicted state set}
\begin{equation}
    \label{eqn:predicted_state_set}
    \hat{\mathcal{X}}_t^i = F_t^i \mathcal{\bar{X}}_{t-1}^i \oplus \mathcal{X}_{Q,t}^i .
\end{equation}
If there exists a measurement set~$\mathcal{X}_t^i$, we then define the \emph{estimated set} of the $i^{th}$ sensor to be
\begin{equation}
    \label{eqn:estimated_set}
    \mathcal{\bar{X}}_t^i = \hat{\mathcal{X}}_t^i \cap \mathcal{X}_t^i. 
\end{equation}

A confidence metric $c_t^i$ is then assigned to the estimated set~$\mathcal{\bar{X}}_t^i$. We define the confidence to be the volumetric \emph{intersection over union}
of the predicted set and estimated set for the current time step, i.e., 
\begin{equation}
    \label{eqn:confidence_IoU}
    c^i_t = \frac{\vol{\bar{\mathcal{X}}^i_t \cap \mathcal{\hat{X}}^i_t}}{\vol{\mathcal{\bar{X}}^i_t \cup \mathcal{\hat{X}}^i_t}} = \frac{\vol{\bar{\mathcal{X}}_t^i}}{\vol{\hat{\mathcal{X}}_t^i}}\;,
\end{equation}
where both the numerator and denominator are simplified using the fact that $\mathcal{\bar{X}}_t^i = \bar{\mathcal{X}}_t^i \cap \mathcal{\hat{X}}_t^i \subseteq \hat{\mathcal{X}}_t^i$. The intersection over union provides a metric for the overall similarity between two sets; when the sets have a large overlap then the metric approaches the value one, and as the sets overlap less then the metric approaches zero. This metric is used in the computer vision and object detection literature~\citep{Rezatofighi2019_IoU}, but to the authors' knowledge, it has not been used for set-based estimation, and the proposed methods combine this metric with hybrid zonotopes in a novel manner. 
If a measurement set~$\mathcal{X}_t^i$ does not exist, for example due to packet loss or when the V2X unit no longer detects the pedestrian, then the predicted set is used as the estimated set.

One challenge with using volumetric intersection over union as a confidence metric is the exponential computational and time complexity it inherits form the constrained zonotope vertex enumeration problem. This poses particular challenges in an estimation framework, where the number of generators tends to compound from one time step to the next as new measured sets are intersected with the predicted sets. This can be mitigated in part by utilizing the latest algorithms to simplify constrained zonotope representations~\cite{raghuraman2022set, Kopetzki2017}. Further, the following fusion technique does not rely on the use of this confidence metric specifically, and the metric could be exchanged for another in use cases when the vertex enumeration is not possible.

\subsubsection{Fusion Center}
Given the estimated sets $\{ \mathcal{\bar{X}}_t^1, \dots, \mathcal{\bar{X}}_t^n \}$ from $n$ different sensors, as well as their confidences $\{ c_t^1, \dots, c_t^n \}$, the fused set~$\confSet$ is constructed as
\begin{equation}
\label{eqn:combine_confidence_confSet_defn}
    \confSet = \bigcup_{\varnothing \neq \mathcal{S} \subseteq \mathcal{N}} \left( \bigcap_{i\in \mathcal{S}} \mathcal{\bar{X}}_t^i \times \left\{ \frac{1}{n}\sum_{i\in\mathcal{S}} c_t^i \right\}\right),
\end{equation}
where $\mathcal{N}=\{1,\dots,n\}$. The union in~\eqref{eqn:combine_confidence_confSet_defn} is performed over all sets of indices $\mathcal{S}$ where $\varnothing \neq \mathcal{S}\subseteq\mathcal{N}$. The set $\confSet$ can be used to examine how well the estimated sets overlap, since the confidences add together for states that lie in the intersection of multiple estimated sets. The set $\confSet$ exists in dimension $\gamma+1$, where the new dimension sums together the confidences of all estimated sets that contain a given point (and normalizes by dividing by the number of sensors).

Rather than combinatorial looping through all possible subsets $\mathcal{S}\subseteq\mathcal{N}$ and the binary complexity inherited through so many unions, $\confSet$ is constructed using Algorithm~\ref{alg:construct_confSet}. Algorithm~\ref{alg:construct_confSet} takes as input the estimated sets and confidences, along with a ``feasible space" $\mathcal{F}_t$ that bounds all of the measurements. It is important that all of the measured sets be contained within $\mathcal{F}_t$ so that they are fully included in the fused set. One could use $\mathcal{F}_t = \cup _{i \in \mathcal{N}} \bar{\mathcal{X}}_t^i$, though this adds unnecessary complexity in terms of the number of binary generators. A more efficient approach is to make $\mathcal{F}_t$ a giant box, centered at the detection area and represented as a $\gamma$-dimensional zonotope, large enough to cover the entire area where measurements could be conceived. Line 4 of Algorithm~\ref{alg:construct_confSet} appends each estimated set with an additional dimension for its confidence, and sets the confidence to zero at all points that are not in the set. Line 7 sets up $\confSet$ to have the desired structure of summing along the confidence dimensions. Line 9 enforces the actual estimated sets and confidences upon $\confSet$, and Line 11 projects away the dimensions corresponding to the confidence of each set individually. Line 11 also scales the final dimension of $\confSet$ (representing the summed confidences over intersections) by $\frac{1}{n}$, so that this dimension acts as an average of the confidences for estimated sets that include a given point in the measurement space.

\begin{algorithm}
    \caption{Fusion center at time $t$}
    \begin{algorithmic}[1]
    \State \textbf{Input:} $\{\bar{\mathcal{X}}^1_t, \dots, \bar{\mathcal{X}}^n_t\}$ where $\bar{\mathcal{X}}^i_t \subset \Rspace^\gamma\ \forall i\in \mathcal{N}$, $\mathcal{N}=\{1,\dots,n\}$, $\{c^1_t,\dots,c^n_t\}$, $\mathcal{F}_t\supseteq \cup_{i\in\mathcal{N}} \bar{\mathcal{X}}^i_t$, and $\mathcal{I} = \interval{0}{1}$
    \State  \textbf{Output:} $\confSet$
        \For{$i\in\mathcal{N}$}
            \State $\mathcal{\Tilde{H}}^i = \left(\bar{\mathcal{X}}^i_t \times \{c^i_t\} \right) \cup \left(\mathcal{F}_t \times \{0\} \right)$
            \State $R^i = \begin{bmatrix}
                \idmatrix_\gamma & \zeros_{\gamma\times n} & \zeros_{\gamma\times 1} \\
                \zeros_{1\times \gamma} & \unitvector{i}{n} & \zeros 
            \end{bmatrix}$
        \EndFor
        \State $\mathcal{H}_\text{temp} \leftarrow \begin{bmatrix}
            \idmatrix_{\gamma+n} \\ \zeros_{1\times \gamma} \ \ \ones_{1 \times n}
        \end{bmatrix} \left(\mathcal{F}_t \times \underbrace{\mathcal{I}  \times\cdots \times \mathcal{I}}_n\right)$
        \For{$i\in\mathcal{N}$}
            \State $\mathcal{H}_\text{temp} \leftarrow \mathcal{H}_\text{temp} \cap_{R^i} \mathcal{\Tilde{H}}^i$
        \EndFor
        \State $\confSet = \begin{bmatrix}
            \idmatrix_\gamma & \zeros_{\gamma \times n} & \zeros_{\gamma \times 1} \\ \zeros_{1\times \gamma} & \zeros_{1 \times n} & 1/n 
        \end{bmatrix} \mathcal{H}_\text{temp}$
    \end{algorithmic}
    \label{alg:construct_confSet}
\end{algorithm}

If the estimated set $\bar{\mathcal{X}}_t^i$ (where $i\in\mathcal{N}$) is a constrained zonotope with $e^i$ generators and $\gamma_c^i$ constraints, and assuming $\mathcal{F}_t$ is a zonotope with $\gamma$ generators (the smallest it could be), then $\confSet$ as constructed in Algorithm~\ref{alg:construct_confSet} has $e$ continuous generators, $p$ binary generators, and $\gamma_c$ constraints, where
\begin{subequations}
    \begin{gather}
        e = (3+\gamma)n + \gamma + \sum_{i=1}^n e^i\;,\\
        p = 2n\;,\\
        \gamma_c = (4+\gamma)n+\sum_{i=1}^n \gamma_c^i\;.
    \end{gather}
\end{subequations}%

Using the combined confidence set $\confSet$, we can define a function
\begin{equation}
    \label{eqn:conf_func}
    C(x) = \max_{(x, c)\in \confSet} c\;,
\end{equation}
where $C(x)$ is the maximum confidence given a state $x\in \Rspace^\gamma$. Representing this function in practice may be difficult and computationally costly, as it involves considering each possible intersection of measurements, the exact combinatorial problem that was being avoided in the construction of $\confSet$. However, similar behavior can be achieved by instead examining the maximum confidence over a region of interest $\mathcal{P}$, which could be the EV reachable set ~$\mathcal{R}_{[t,t+\Delta \tau]}$ or another set such as a pedestrian crossing or detected obstacle. The maximum confidence over this region can be found by performing the generalized intersection 
\begin{equation}
    \mathcal{H}_\mathcal{P} = \confSet \cap_{[\idmatrix \ \zeros]} \mathcal{P}\;,
\end{equation}
and then computing the support function 
\begin{equation}
    \label{eqn:max_conf_over_P}
    c_\text{max} = \max_{z\in \mathcal{H}_\mathcal{P}} [\zeros_{1\times\gamma}\ 1] z\;.
\end{equation}
Alternatively, one could also compute a halfspace intersection of $\confSet$ along the final dimension and project back onto the measurement space to determine a hybrid zonotope set of all measured points that are above some confidence threshold.

The proposed fusion method uses hybrid zonotopes to represent unions of other zonotopic sets; however, certain queries of the fused set, such as~\eqref{eqn:conf_func} and \eqref{eqn:max_conf_over_P}, now require mixed-integer optimization, which brings additional computation complexity to the implementation of this method. The optimization can be performed using general-purpose commercial solvers such as Gurobi~\citep{gurobi}, or using the open source toolbox ZonoOpt~\citep{robbins2026_hybrid} with a tailored mixed-integer optimization routine that takes advantage of the specific structure of hybrid zonotopes and can be implemented on embedded hardware. 
\section{Simulation}
\label{sec:simulation_setup_results}

\newcommand{\subfigwidth}{0.325\textwidth}
\newcommand{\subfigvspace}{1em}
\begin{figure*}
    \centering
    \begin{subfigure}{\subfigwidth}
        \captionsetup{skip=0pt}
        \includegraphics[width=\linewidth, trim = .2in .7in .6in 1.2in, clip]{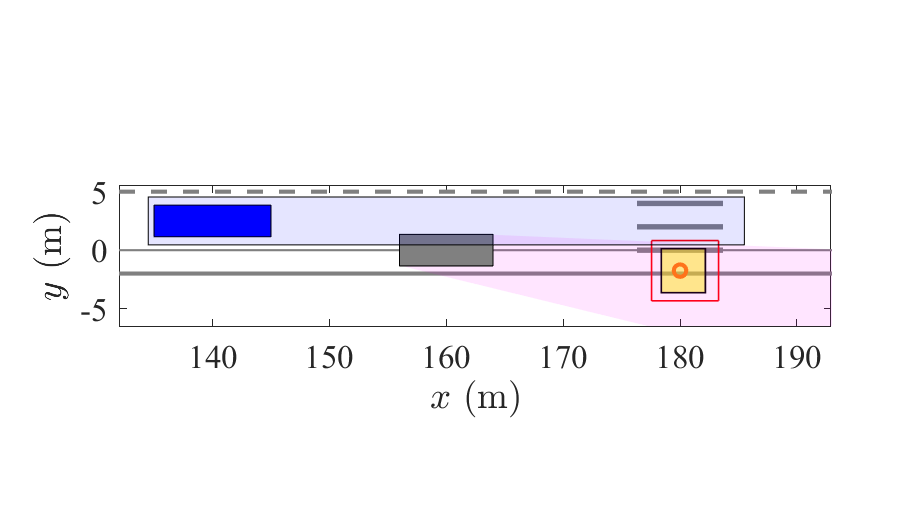}
        \caption{Case 1: $c_\text{cross} = 0.80$, $c_\text{reach} = 0.28$, $v=20$}
        \label{subfig:nonoise_a}
    \end{subfigure} \hfill
    \begin{subfigure}{\subfigwidth}
        \captionsetup{skip=0pt}
        \includegraphics[width=\linewidth, trim = .2in .7in .6in 1.2in, clip]{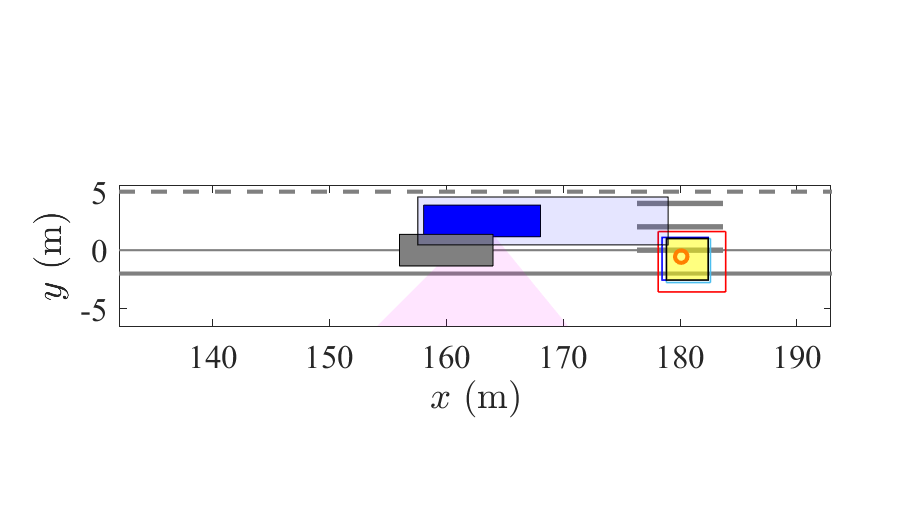}
        \caption{Case 1: $c_\text{cross} = 0.81$, $c_\text{reach} = 0.81$, $v=5.2$}
        \label{subfig:nonoise_b}
    \end{subfigure} \hfill
    \begin{subfigure}{\subfigwidth}
        \captionsetup{skip=0pt}
        \includegraphics[width=\linewidth, trim = .2in .7in .6in 1.2in, clip]{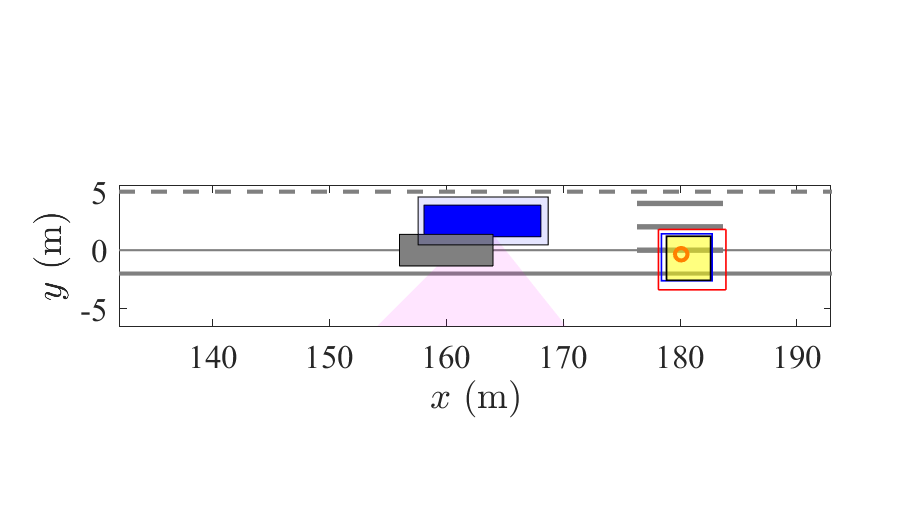}
        \caption{Case 1: $c_\text{cross} = 0.81$, $c_\text{reach} = 0$, $v=0$}
        \label{subfig:nonoise_c}
    \end{subfigure}
    \\
    \vspace{\subfigvspace}
    \begin{subfigure}{\subfigwidth}
        \captionsetup{skip=0pt}
        \includegraphics[width=\linewidth, trim = .2in .7in .6in 1.2in, clip]{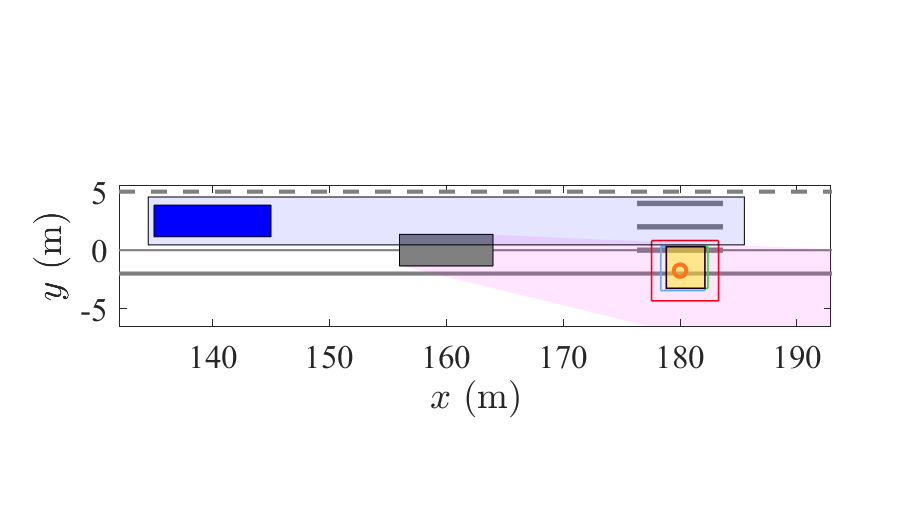}
        \caption{Case 2: $c_\text{cross} = 0.74$, $c_\text{reach} = 0.28$, $v=20$}
        \label{subfig:noisy_a}
    \end{subfigure} \hfill
    \begin{subfigure}{\subfigwidth}
        \captionsetup{skip=0pt}
        \includegraphics[width=\linewidth, trim = .2in .7in .6in 1.2in, clip]{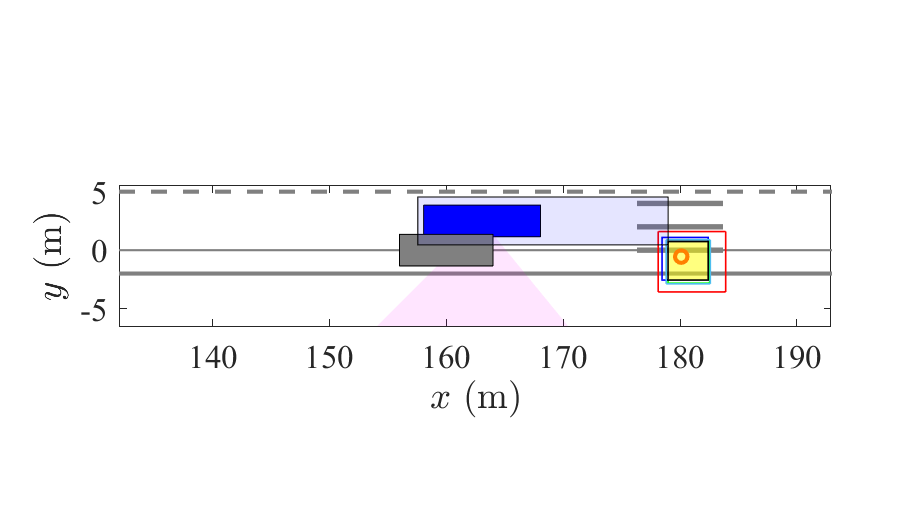}
        \caption{Case 2: $c_\text{cross} = 0.76$, $c_\text{reach} = 0.62$, $v=5.2$}
        \label{subfig:noisy_b}
    \end{subfigure} \hfill
    \begin{subfigure}{\subfigwidth}
        \captionsetup{skip=0pt}
        \includegraphics[width=\linewidth, trim = .2in .7in .6in 1.2in, clip]{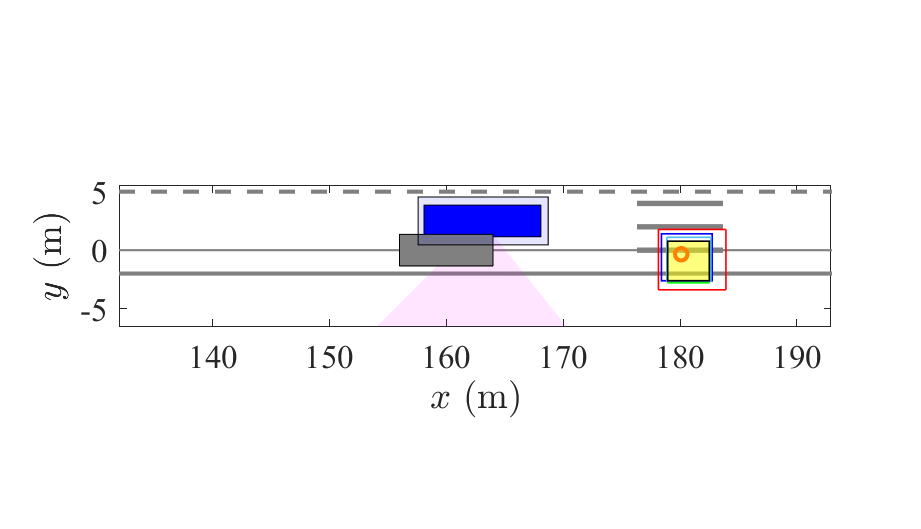}
        \caption{Case 2: $c_\text{cross} = 0.77$, $c_\text{reach} = 0$, $v=0$}
        \label{subfig:noisy_c}
    \end{subfigure}
    \\
    \vspace{\subfigvspace}
    \begin{subfigure}{\subfigwidth}
        \captionsetup{skip=0pt}
        \includegraphics[width=\linewidth, trim = .2in .7in .6in 1.2in, clip]{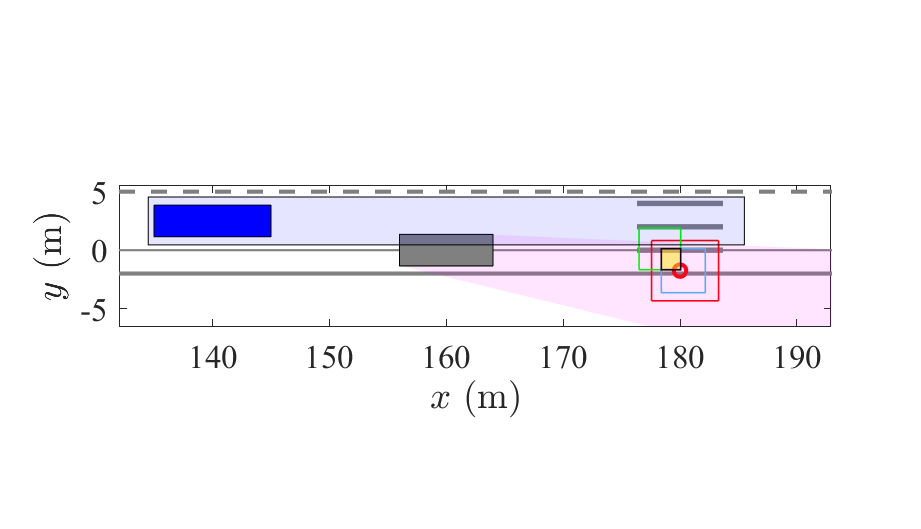}
        \caption{Case 3: $c_\text{cross} = 0.31$, $c_\text{reach} = 0$, $v=14$}
        \label{subfig:offset_a}
    \end{subfigure} \hfill
    \begin{subfigure}{\subfigwidth}
        \captionsetup{skip=0pt}
        \includegraphics[width=\linewidth, trim = .2in .7in .6in 1.2in, clip]{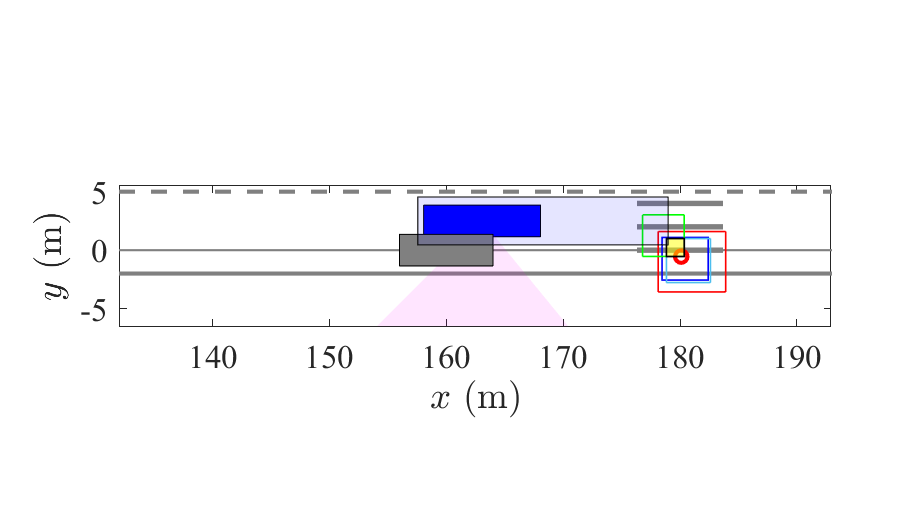}
        \caption{Case 3: $c_\text{cross} = 0.62$, $c_\text{reach} = 0$, $v=4$}
        \label{subfig:offset_b}
    \end{subfigure} \hfill
    \begin{subfigure}{\subfigwidth}
        \captionsetup{skip=0pt}
        \includegraphics[width=\linewidth, trim = .2in .7in .6in 1.2in, clip]{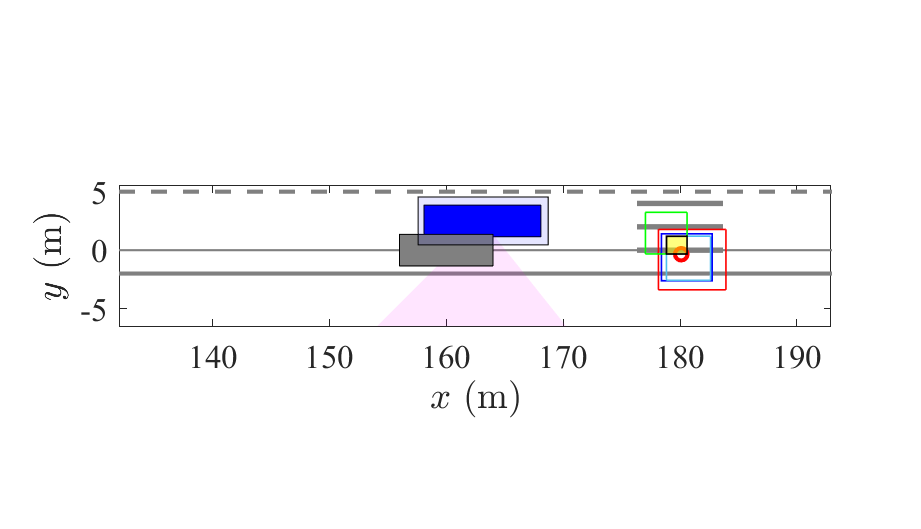}
        \caption{Case 3: $c_\text{cross} = 0.70$, $c_\text{reach} = 0$, $v=0$}
        \label{subfig:offset_c}
    \end{subfigure}
    \caption{Each row presents results of the MATLAB simulation for one of three cases. In case 1, the RSUs have no additive noise. In case 2, the RSUs have random additive noise. In case 3, one of the RSUs has a significant offset. The columns show different time instances of the simulation, progressing from left to right. In all three case studies, despite the noise and even disjunction in sensor data, the EV comes to a stop due to a sufficiently large confidence of the pedestrian being in the crosswalk.}
    \label{fig:MATLAB_sim_results}
\end{figure*}

A simulation of the proposed methodology was performed using MATLAB on a Windows 11 desktop with an Intel\textregistered~Core\textsuperscript{\tiny TM} i7-14700 × 28 processor and 32~GB of RAM. The simulated scenario is depicted by Fig.~\ref{fig:MATLAB_sim_overview}. 
This work uses the zonoLAB toolbox~\citep{koeln2024zonolab} for storing and representing zonotopic sets in MATLAB. The combined confidence set arising from those three measurements is used to determine the EV's control action. 

The ego controller is given the fused set $\confSet$, a reachable set of the ego vehicle $\mathcal{R}_{[t, t+T]}$ over prediction time $T=2$s assuming constant current speed $v_k$, and the crosswalk location as a set $\mathcal{X}_C$ along with the current distance to the crosswalk $d_{\text{cross}}$. The distance of the reachable set in front of the vehicle is $d_{\text{reach}} = vT$. Define the intersections $\mathcal{X}_{\text{cross}} = \confSet \cap_{[\idmatrix\ \zeros]} \mathcal{X}_C$ and $\mathcal{X}_{\text{reach}} = \confSet \cap_{[\idmatrix\ \zeros]} \mathcal{R}_{[t,t+T]}$ to represent the intersection of the combined confidence set with the crosswalk and with the ego vehicles reachable set, respectively. Then, the maximum confidences over those regions can be found via the support function calls $c_{\text{cross}} = \max_{x\in \mathcal{X}_{\text{cross}}} [0\ 0\ 1] x$ and $c_{\text{reach}} = \max_{x\in \mathcal{X}_{\text{reach}}} [0\ 0\ 1] x$. The controller selects the new speed $v_{k+1}$ to be
\begin{equation}
    \label{eqn:ego_control_law}
    \begin{aligned}
        v_{k+1} = \min\Bigg\{ &\frac{v_k}{1+e^{-k(c_\text{reach}) (d_\text{reach}-d(c_\text{reach}))}}\;, \\
        &\frac{v_k}{1+e^{-k(c_\text{cross})(d_\text{cross}-d(c_\text{cross}))}}\Bigg\}\;,
    \end{aligned}
\end{equation}%
where
\begin{equation}
    \label{eqn:ego_control_params}
    \begin{aligned}
        k(c)&=\begin{cases}
            0.3 & \text{if } 0\leq c < 0.5\;,\\
            0.6 & \text{if } 0.5\leq c < 0.8\;,\\
            1 &\text{otherwise}\;,
        \end{cases}\\
        d(c)&= \begin{cases}
            5 & \text{if } 0\leq c < 0.5\;,\\
            10 & \text{if } 0.5\leq c < 0.8\;,\\
            15 &\text{otherwise}\;.
        \end{cases}
    \end{aligned}
\end{equation}%
Sigmoid functions are used in~\eqref{eqn:ego_control_law} to scale the speed according to the distance to the pedestrian or crossing (whichever requires harsher control action), where the sigmoid parameters are determined in~\eqref{eqn:ego_control_params} based on logic thresholds of the confidence. Note that this is only an example of how the combined confidence hybrid zonotope can be incorporated into a control law and does not provide any rigorous guarantee of safety. 

\begin{figure}[bh!]
    \centering
    \includegraphics[trim={0.35cm 0.1cm 0.43cm 0.5cm},clip,width=1\linewidth]{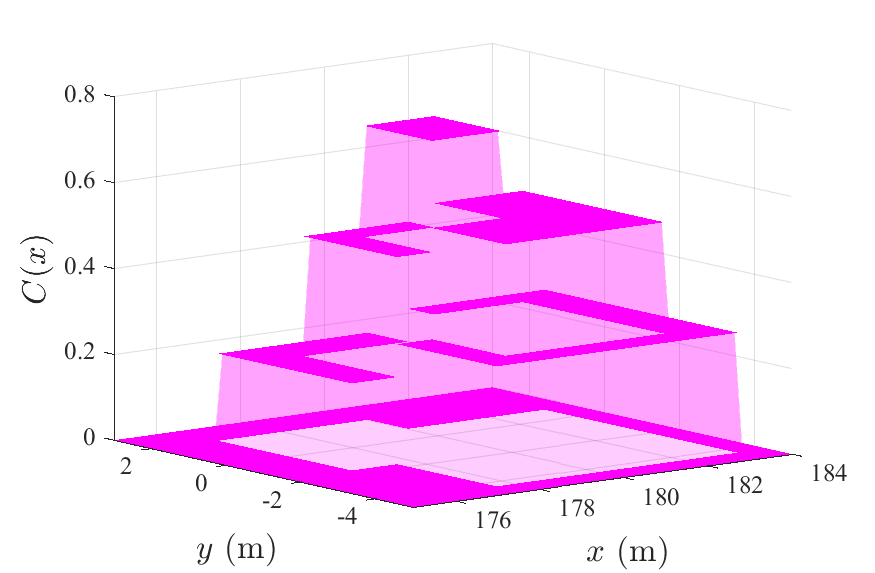}
    \caption{Plot of the confidence function $C(x)$ from~\eqref{eqn:conf_func} for the simulated scenario depicted in Fig.~\ref{subfig:offset_a}.}
    \label{fig:matlab_sim_conf_func}
\end{figure}

\subsection{Results}
We consider three cases: Case 1, where there is no noise in the measurements from the RSUs, case 2, where there is random additive noise to the measurements, and case 3, where one of the RSUs experiences a constant bias (along with small random noise on both RSUs). Each row of Fig.~\ref{fig:MATLAB_sim_results} depicts one of these cases, with columns showing different time instances. The plots depict the location of all vehicles along with estimated sets of the pedestrians location. In all three cases, the EV slows its speed as it approaches the pedestrian crossing and comes to a stop at a safe distance in front of the crossing.
Even in case 3 where the intersection is empty or very small, the EV still takes appropriate control action if the fused confidence of a pedestrian in the crossing or the reachable set exceeds the threshold defined in the controller. The exact values of the confidence and the speed of the EV in the first and second columns of Fig.~\ref{fig:MATLAB_sim_results} vary accordingly between the different case studies, but in all cases, the vehicle comes to a stop before the crossing. Across all three cases with 100 time steps each, the average computation time for a single time step (estimation and fusion) was 0.281s, with a maximum time of 0.370s. This is greater than the simulated time step of 0.1s, however, as shown in the next section, implementing the proposed approach in C++ rather than MATLAB can achieve real-time execution.

Fig.~\ref{fig:matlab_sim_conf_func} plots the confidence function $C(x)$ as defined in~\eqref{eqn:conf_func} for the case and time instance shown in Fig.~\ref{subfig:offset_a}. This is similar to the fused set $\confSet$, but different in that it only considers the maximum combined confidence associated with each point in the state space. Representing this function as a set requires taking complements of the estimated sets, which can be done using the identity in~\cite{bird2021unions}, but has a high computational complexity. Therefore the controller formulation only uses the fused set $\confSet$, but the combined confidence function is shown here for visualization.

\begin{figure}
    \centering
    \includegraphics[width=0.9\linewidth]{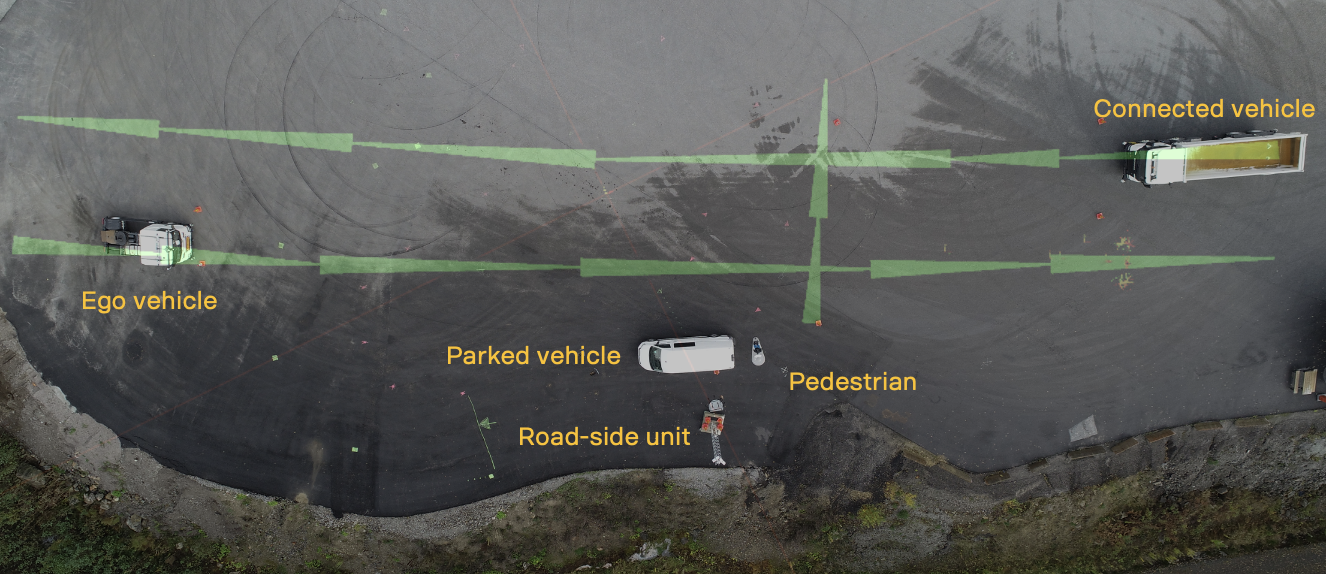}
    \caption{Aerial view of the experimental scenario with three vehicles, one RSU, and a pedestrian. The pedestrian is in the occluded area of the ego vehicle. This picture is taken at Scania's test track.}
    \label{fig:real-exp-setup}
\end{figure}

\section{Experiment}
\label{sec:real_setup_results}
This section first discusses the experimental setup, followed by results and evaluation of the proposed framework.
Fig.~\ref{fig:real-exp-setup} provides an aerial view of the Scania test track with the occluded pedestrian crossing scenario. The truck located on the left acts as the EV and the van parked in front occluding the pedestrian in the EV's FOV is the PV. A truck on the right acts as the CV, which is able to share perception data to the EV. A camera, acting as the RSU, is mounted on a $7$ meters tall pole and can share perception data to the EV. The high-definition (HD) map with two lanes and a pedestrian-crossing is superimposed on the image. The green arrows indicate the direction of each lane. 

The proposed method was implemented in C++ using the ZonoOpt library~\citep{robbins2025sparsity_zonoopt} on the Traton Autonomous Platform, deployed on Scania's autonomous prototype vehicles. These vehicles are equipped with technologies and an HD-map for self localization, perception, motion planning, and control. The perception layer generates object detections using sensor configurations mounted on the vehicles. The vehicles are equipped with V2X modules that share the detection wrapped in collaborative perception messages~(CPMs) according to standards detailed in~\cite{etsi_cpm}. The messages are transmitted over dedicated short-range communication~(DSRC) built on ITS-G5 standards~\citep{etsi_its_g5}. These messages are transmitted at a rate of $10~\mathrm{Hz}$. We used a camera from Viscando\footnote{\url{https://viscando.com/}} which is equipped with a perception module that processes camera images\footnote{Images are processed directly after capture and permanently removed within $20\mathrm{ms}$ after being captured. Only anonymized trajectories of perceived road-users are stored and transmitted. This ensures full GDPR~(general data protection regulation) compliance and possibility to deploy Viscando sensors on high-security areas like test tracks and on public roads.} to produce object detections. These detections are then shared with the ego vehicle over a 5G network. The detections are sent at a rate of $20~\mathrm{Hz}$. The sensor measurements from each V2X unit and EV is recorded using this setup. This recorded log is then replayed along with simulation of the EV and internal visualization tool. 

\subsection{Results}
\begin{figure*}
    \centering
    \begin{subfigure}{.4\linewidth}
        \includegraphics[width=1.\linewidth]{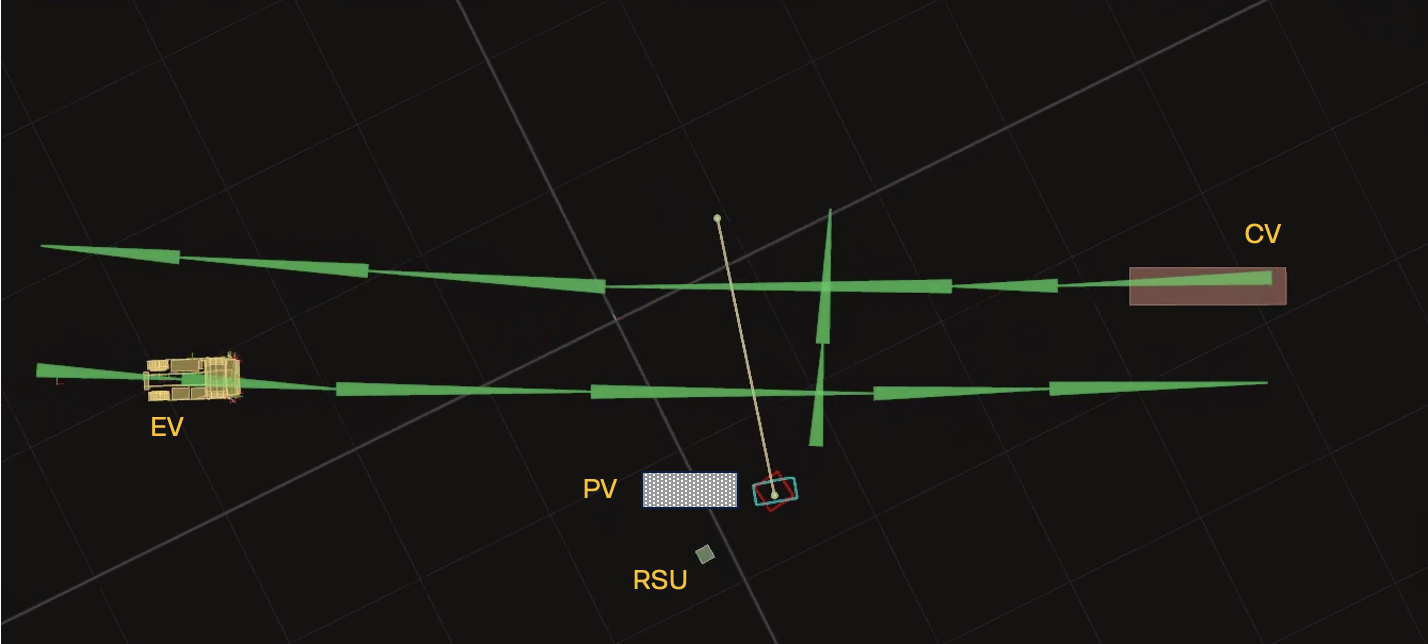}
        \caption{At $t=19.74~\mathrm{s}$, two measurement sets,~$\mathcal{X}^1_t$ and $\mathcal{X}^2_t$, are obtained from the RSU and CV, respectively.}
        \label{fig:6a}
    \end{subfigure} \hfill
        \begin{subfigure}{.29\linewidth}
        \includegraphics[width=1\linewidth]{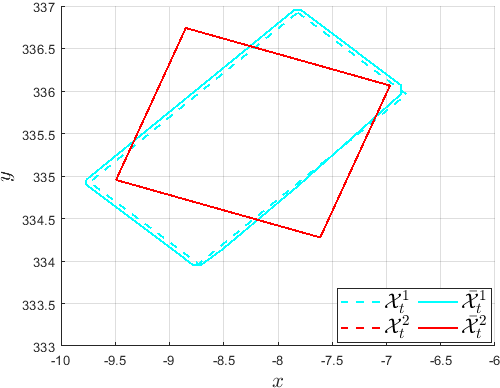}
        \caption{}
        \label{fig:6b}
    \end{subfigure} \hfill
        \begin{subfigure}{.29\linewidth}
        \includegraphics[width=1\linewidth]{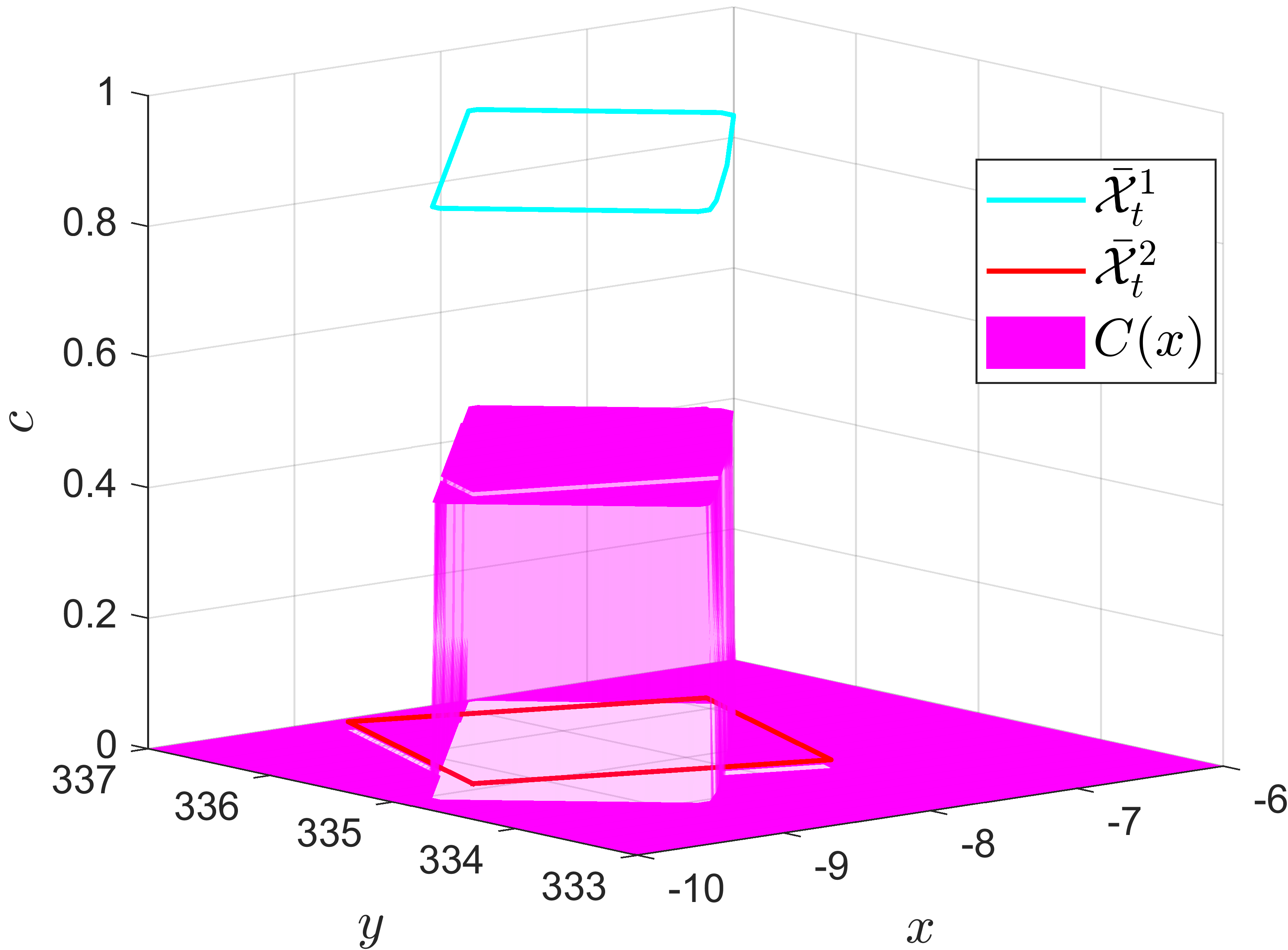}
        \caption{}
        \label{fig:6c}
    \end{subfigure} \hfill
        \begin{subfigure}{.4\linewidth}
        \centering
        \includegraphics[width=1.\linewidth]{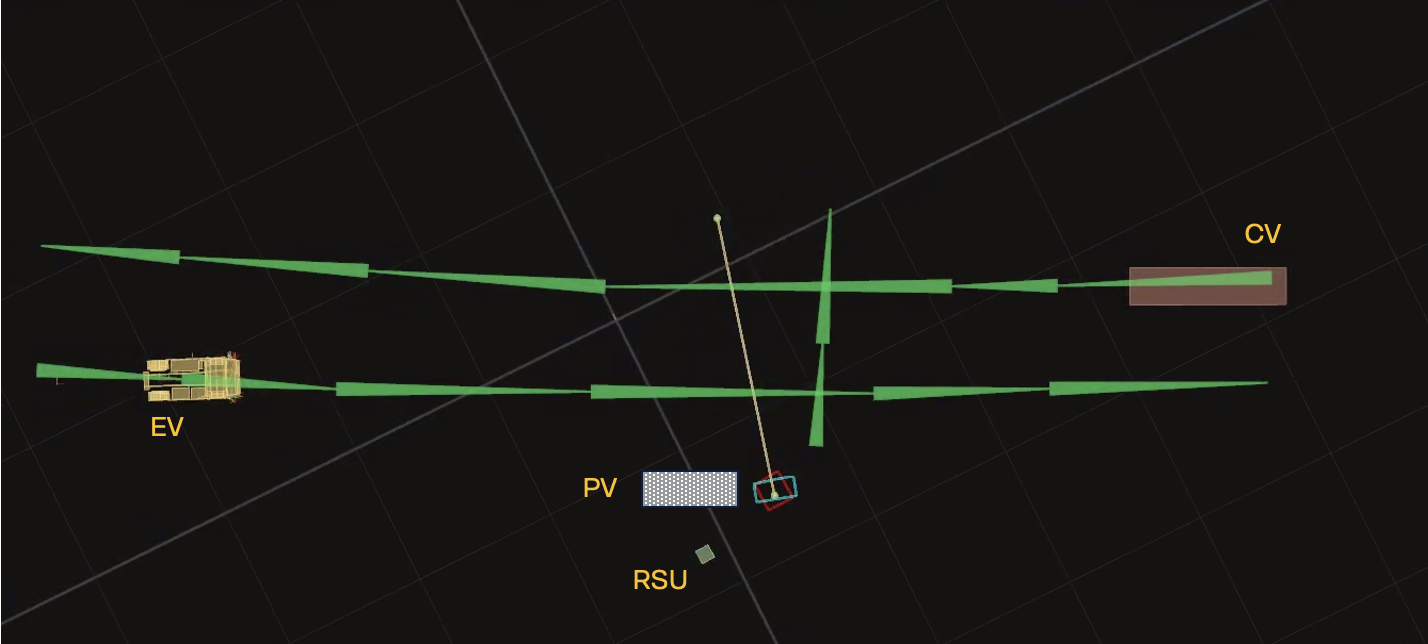}
        \caption{At $t=19.79~\mathrm{s}$, two measurement sets,~$\mathcal{X}^1_t$ and $\mathcal{X}^2_t$, are obtained from the RSU and CV, respectively.}
        \label{fig:6d}
    \end{subfigure} \hfill
        \begin{subfigure}{.29\linewidth}
        \includegraphics[width=1\linewidth]{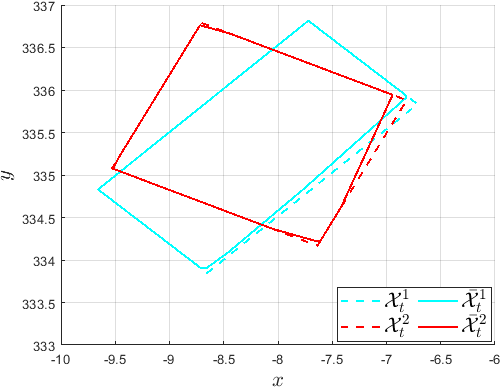}
        \caption{}
        \label{fig:6e}
    \end{subfigure} \hfill
        \begin{subfigure}{.29\linewidth}
        \includegraphics[width=1\linewidth]{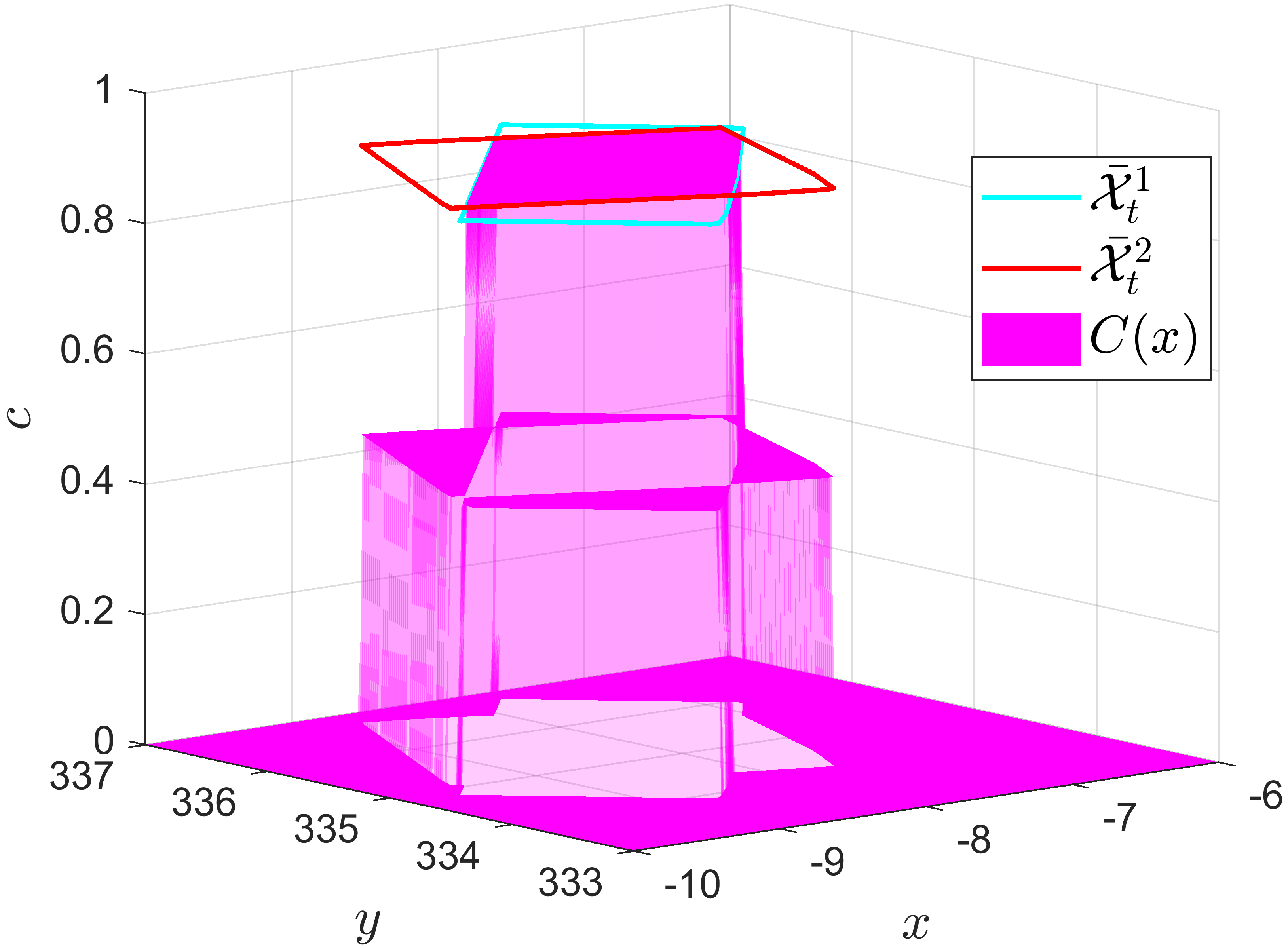}
        \caption{}
        \label{fig:6f}
    \end{subfigure} \hfill
        \begin{subfigure}{.4\linewidth}
        \includegraphics[width=1.\linewidth]{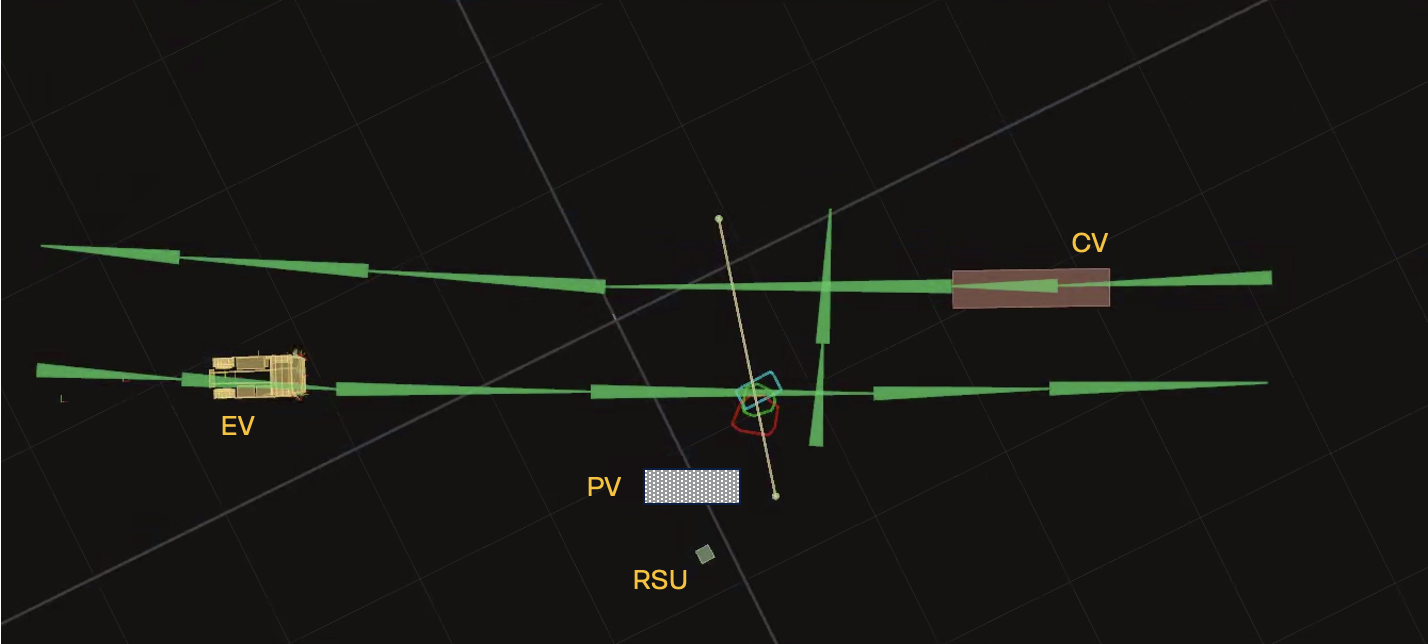}
        \caption{At $t=23.06~\mathrm{s}$, three measurement sets, $\mathcal{X}^1_t$,~$\mathcal{X}^2_t$, and $\mathcal{X}^3_t$, are obtained from the RSU, CV, and EV, respectively.}
        \label{fig:6g}
    \end{subfigure} \hfill
        \begin{subfigure}{.29\linewidth}
        \includegraphics[width=1\linewidth]{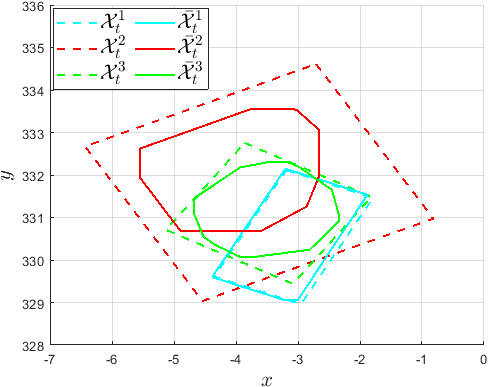}
        \caption{}
        \label{fig:6h}
    \end{subfigure} \hfill
        \begin{subfigure}{.29\linewidth}
        \includegraphics[width=1\linewidth]{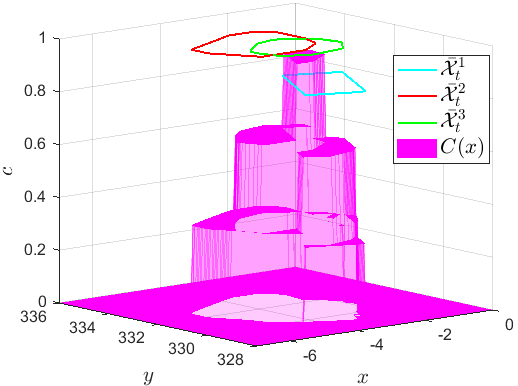}
        \caption{}
        \label{fig:6i}
    \end{subfigure} \hfill
    \caption{The first column presents snapshots from the internal visualization tool of the EV at three time instances, with the yellow line representing the pedestrian ground truth. The second column shows the measurement and estimated sets at each time instance. The third column shows the estimated sets and confidence function $C(x)$ at each time instance.}
    \label{fig:three_instances_of_real_experiment}
\end{figure*}

Fig.~\ref{fig:three_instances_of_real_experiment} presents snapshots of the internal visualization tool of the EV with results from the proposed method at three time instances. In each subfigure, $\bar{\mathcal{X}}^1_t$ denotes the estimated set computed based on the measurement set $\mathcal{X}^1_t$ obtained from the RSU. Similarly, $\bar{\mathcal{X}}^2_t$ and $\bar{\mathcal{X}}^3_t$ denote the estimated sets based on the measurement sets obtained from the CV and the EV, respectively. The estimated sets $\mathcal{\bar{X}}^1_t$, $\mathcal{\bar{X}}^2_t$, and $\mathcal{\bar{X}}^3_t$ are represented by blue, red, and green constrained zonotopes, respectively. $C(x)$ denotes the confidence function of the fused set~$\confSet$ computed using the given estimated sets and is represented by the magenta region. Fig.~\ref{fig:6a} present the scene at $t=19.74~\mathrm{s}$, when the EV receives measurement sets from the RSU and CV. There is no measurement set from the EV because the pedestrian is occluded from the EV's FOV. The measurement and estimated sets at this time instance are presented in Fig.~\ref{fig:6b}. Fig.~\ref{fig:6c} presents the confidence metric of these estimated sets and the confidence region of the fused set. 
The RSU has been sending measurement sets for the detected pedestrian for a few seconds before this time instance, therefore, the confidence~$c^1_t = 0.90$. The measurement set from the CV is relatively new, and its predicted region is larger than the measurement set. The intersection of these sets leads to a smaller estimated set, resulting in a confidence metric of~$c^2_t = 0.01$. This results in a maximum confidence of the fused set of~$0.45$.

Fig.~\ref{fig:6d} presents a slightly later time instance where the RSU and the CV continue to detect the pedestrian. It can be observed in Fig.~\ref{fig:6e} that the estimated sets are slightly tighter than the estimated sets in Fig.~\ref{fig:6b}. Both estimated sets in Fig.~\ref{fig:6e} have high confidence metrics, with $c^1_t = 0.88$ and~$c^2_t = 0.88$. This results in a maximum confidence of the fused set of~$0.88$. Fig.~\ref{fig:6g} presents the scene at time instance~$t=23.06~\mathrm{s}$ when the pedestrian starts to cross the road and the EV is also able to detect the pedestrian. Fig.~\ref{fig:6h} presents the three estimated sets from the RSU, CV, and EV. The estimated set $\mathcal{\bar{X}}^1_t$ has confidence~$c^1_t = 0.86$. Both the estimated sets from the CV and the EV are smaller than the measurement set, meaning that the intersection between the predicted set and the measurement set is equal to the predicted set, resulting in confidences~$c^2_t = 1.0$ and~$c^3_t = 1.0$. This results in a maximum confidence of the fused set of~$0.95$. During the experiment, the computation time averaged $0.094\mathrm{s}$ per time step.

\begin{figure*}[t]
    \centering
    \begin{subfigure}{.24\linewidth}
        \includegraphics[width=1\linewidth]{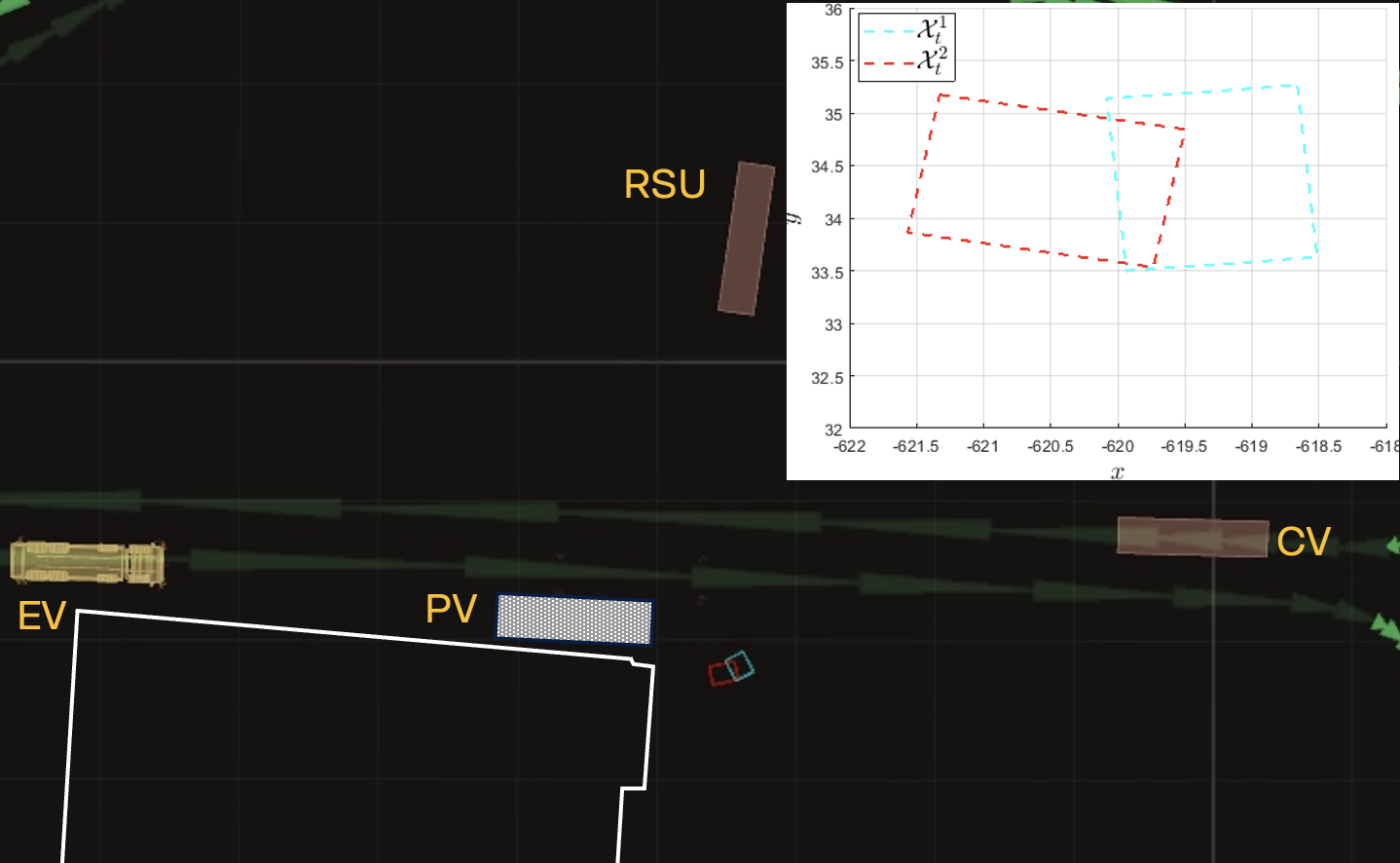}
    \caption{At $t=3.08~\mathrm{s}$, two measurement sets, $\mathcal{X}^1_t$ and $\mathcal{X}^2_t$, are obtained from the RSU and CV, respectively.}
    \label{fig:7a}
    \end{subfigure}
    \hfill
    \begin{subfigure}{.24\linewidth}
        \includegraphics[width=1\linewidth]{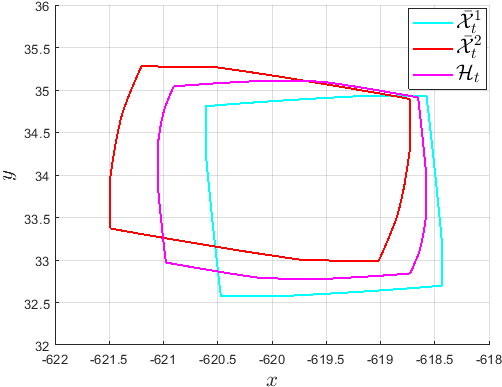}
        \caption{Using zonotopes.}
        \label{fig:7b}
    \end{subfigure}
    \hfill
    \begin{subfigure}{.24\linewidth}
        \includegraphics[width=1\linewidth]{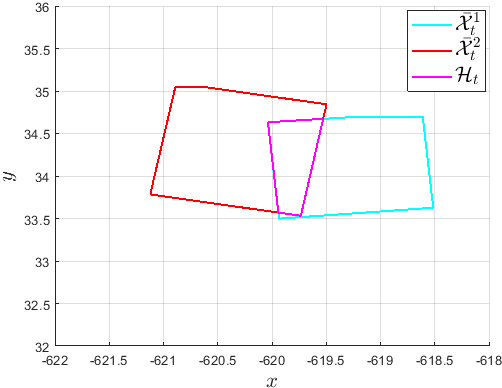}
        \caption{Using constrained zonotopes.}
        \label{fig:7c}
    \end{subfigure}
    \hfill
    \begin{subfigure}{.24\linewidth}
        \includegraphics[width=1\linewidth]{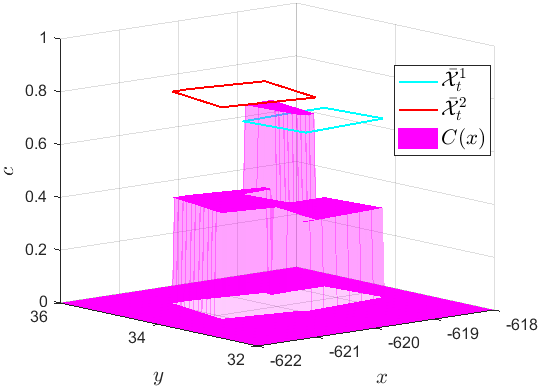}
        \caption{Using constrained zonotopes and hybrid zonotopes.}
        \label{fig:7d}
    \end{subfigure}
    \caption{Comparison between set-based estimation and fusion using different set representations.}
    \label{fig:comparision_zono_cZ_HZ}
\end{figure*}

In Fig.~\ref{fig:comparision_zono_cZ_HZ}, we consider a time instance from a similar scenario on the Scania test track. At this time instance, the overlap between the measurement sets from the two V2X units is below $50\%$ of area. We compare the results using three frameworks: (1) Set-based estimation and fusion using zonotopes~\citep{narri2023shared}, (2) set-based estimation and fusion using constrained zonotopes~\citep{narri2025situational}, and (3) set-based estimation and fusion using constrained zonotopes and hybrid zonotopes (the proposed framework). Fig.~\ref{fig:7a} presents a snapshot of the internal visualization tool of the EV at $t=3.08~\mathrm{s}$ and a MATLAB plot of the measurement sets on the top right corner. Fig.~\ref{fig:7b} presents estimated sets~$\mathcal{\bar{X}}^1_t$ and~$\mathcal{\bar{X}}^2_t$ as blue and red zonotopes, respectively, and the fused set~$\mathcal{H}_t$ is presented as a magenta zonotope. It can be observed the resultant sets are over-approximated and are bigger than the measurement sets due to the property of zonotopes being centrally symmetric. When using constrained zonotopes in Fig.~\ref{fig:7c}, the estimated and fused sets are exact intersections, leading to tighter sets. But if one of the measurement sets is inconsistent with the ground truth, then the fused set might not contain the location of the pedestrian. For the cases with zonotopes and constrained zonotopes only, the fused set is considered to have confidence equal to one within the intersection and confidence equal to zero outside the intersection. When using the proposed framework, see Fig.~\ref{fig:7d}, the estimated sets are represented by constrained zonotopes and have confidence metrics $c^1_t = 0.68$ and~$c^2_t=0.80$. The maximum confidence of the fused is $0.74$. This information of the fused set and its confidence can help the EV plan a safe trajectory.
\section{Conclusion}
\label{sec:conclu}
This work proposes a framework for shared situational awareness that can compute a fused set for the pedestrian even when encountered with inconsistent measurement sets from multiple sensors.
When the measurements are inconsistent, their intersection can become empty, which poses a problem for other set-based estimation techniques in the literature. 
In fusing the estimated sets from multiple sensors, a hybrid zonotope is constructed which combines the confidences of sensors in regions where their estimates overlap. 
The confidence metric is computed by finding the volumes of certain constrained zonotopes, which is computationally expensive. 
Scalability could be improved via approximate methods such as random sampling or approximating the set with a representation that allows for easier volume computation (zonotopes, ellipsoids, etc.). 
The proposed method is tested in MATLAB simulation as well as data obtained from Scania's automated prototype vehicles. 
This framework attempts to combine the best of set-based estimation and quantitative methods, by computing estimated sets that can account for uncertainties while also providing a confidence metric which can be used for decision-making. 
Possible future work is to explore scenarios with multiple pedestrians, as well as best practices for handling communication delays and their impact on the confidence metric.

\bibliography{ref}

@inproceedings{narri2021set,
  title={Set-membership estimation in shared situational awareness for automated vehicles in occluded scenarios},
  author={Narri, Vandana and Alanwar, Amr and M{\aa}rtensson, Jonas and Nor{\'e}n, Christoffer and Dal Col, Laura and Johansson, Karl Henrik},
  booktitle={2021 IEEE Intelligent Vehicles Symposium (IV)},
  pages={385--392},
  year={2021},
  organization={IEEE}
}

@article{narri2023shared,
  title={Shared Situational Awareness with {V2X} Communication and Set-membership Estimation},
  author={Narri, Vandana and Alanwar, Amr and M{\aa}rtensson, Jonas and Nor{\'e}n, Christoffer and Johansson, Karl Henrik},
  journal={arXiv:2302.05224},
  year={2023}
}

@article{narri2025situational,
  title={Situational awareness using set-based estimation and vehicular communication: An occluded pedestrian-crossing scenario},
  author={Narri, Vandana and Alanwar, Amr and M{\aa}rtensson, Jonas and Pettersson, Henrik and Nordin, Fredrik and Johansson, Karl Henrik},
  journal={Communications in Transportation Research},
  volume={5},
  pages={100190},
  year={2025},
  publisher={Elsevier}
}

@article{ScottJosephKConstrainedZonotopes,
author = {Scott, Joseph K. and Raimondo, Davide M. and Marseglia, Giuseppe Roberto and Braatz, Richard D.},
copyright = {2016 Elsevier Ltd},
issn = {0005-1098},
journal = {Automatica},
language = {eng},
pages = {126-136},
publisher = {Elsevier Ltd},
title = {Constrained zonotopes: A new tool for set-based estimation and fault detection},
volume = {69},
year = {2016},
}

@article{hybrid2023,
  title={Hybrid zonotopes: A new set representation for reachability analysis of mixed logical dynamical systems},
  author={Bird, Trevor J and Pangborn, Herschel C and Jain, Neera and Koeln, Justin P},
  journal={Automatica},
  volume={154},
  pages={111107},
  year={2023},
  publisher={Elsevier}
}

@article{bird2021unions,
  title={Unions and complements of hybrid zonotopes},
  author={Bird, Trevor J and Jain, Neera},
  journal={IEEE Control Systems Letters},
  volume={6},
  pages={1778--1783},
  year={2021}
}

@Article{Sensor_Failures_in_Autonomous_Vehicles,
AUTHOR = {Matos, Francisco and Bernardino, Jorge and Durães, João and Cunha, João},
TITLE = {A Survey on Sensor Failures in Autonomous Vehicles: Challenges and Solutions},
JOURNAL = {Sensors},
VOLUME = {24},
YEAR = {2024},
NUMBER = {16},
ARTICLE-NUMBER = {5108},
PubMedID = {39204805},
ISSN = {1424-8220},
}

@article{ngo2023cooperative,
  author={Ngo Hieu and Fang Hua and Wang Honggang},
  journal={IEEE Transactions on Vehicular Technology}, 
  title={Cooperative Perception With {V2V} Communication for Autonomous Vehicles}, 
  year={2023},
  volume={72},
  number={9},
  pages={11122-11131},
}

@inproceedings{yee2018collaborative,
    author={Yee Ryan and Chan Ellick and Cheng Bin and Bansal Gaurav},
    booktitle={IEEE Intelligent Vehicles Symposium}, 
    title={Collaborative Perception for Automated Vehicles Leveraging Vehicle-to-Vehicle Communications}, 
    year={2018},
    volume={},
    number={},
    pages={1099-1106},
}

@article{CP_methods_datasets_challenges,
    author={Yushan Han and Zhang Hui and Li Huifang and Jin Yi and Lang Congyan and Li Yidong},
    journal={IEEE Intelligent Transportation Systems Magazine}, 
    title={Collaborative Perception in Autonomous Driving: Methods, Datasets, and Challenges}, 
    year={2023},
    volume={15},
    number={6},
    pages={131-151},
}

@article{sensor_noise_factors,
  title={Analysis of automotive camera sensor noise factors and impact on object detection},
  author={Li, Boda and Chan, Pak Hung and Baris, Gabriele and Higgins, Matthew D and Donzella, Valentina},
  journal={IEEE Sensors Journal},
  volume={22},
  number={22},
  pages={22210--22219},
  year={2022},
  publisher={IEEE}
}

@article{conf:set_fault1,
    author = {Vicenç Puig},
    title = {Fault diagnosis and fault tolerant control using set-membership approaches: Application to real case studies},
    journal = {International Journal of Applied Mathematics and Computer Science},
    number = {4},
    volume = {20},
    year = {2010},
    pages = {619--635}
}

@article{jaulin2009nonlinear,
    author={Luc Jaulin},
    journal={IEEE Transactions on Robotics}, 
    title={A Nonlinear Set Membership Approach for the Localization and Map Building of Underwater Robots}, 
    year={2009},
    volume={25},
    number={1},
    pages={88-98},
}

@article{franze2015obstacle,
    title = {The obstacle avoidance motion planning problem for autonomous vehicles: A low-demanding receding horizon control scheme},
    journal = {Systems \& Control Letters},
    volume = {77},
    pages = {1-10},
    year = {2015},
    author = {Giuseppe Franzè and Walter Lucia},
}

@article{garcia2020guaranteed,
    author = {Ramon A. García and L. Orihuela and Pablo Millan and Francisco R. Rubio and M.G. Ortega},
    title = {Guaranteed estimation and distributed control of vehicle formations},
    journal = {International Journal of Control},
    volume = {93},
    number = {11},
    pages = {2729--2742},
    year = {2020},
    publisher = {Taylor \& Francis},
}

@inproceedings{conf:setloc,
    author={P. Bouron and D. Meizel and Ph. Bonnifait},
    booktitle={European Control Conference }, 
    title={Set-membership non-linear observers with application to vehicle localisation}, 
    year={2001},
    volume={},
    number={},
    pages={1255-1260},
}

@article{raimondo2016closed,
  title={Closed-loop input design for guaranteed fault diagnosis using set-valued observers},
  author={Raimondo, Davide M and Marseglia, Giuseppe Roberto and Braatz, Richard D and Scott, Joseph K},
  journal={Automatica},
  volume={74},
  pages={107--117},
  year={2016},
  publisher={Elsevier}
}

@article{rego2020guaranteed,
  title={Guaranteed methods based on constrained zonotopes for set-valued state estimation of nonlinear discrete-time systems},
  author={Rego, Brenner S and Raffo, Guilherme V and Scott, Joseph K and Raimondo, Davide M},
  journal={Automatica},
  volume={111},
  pages={108614},
  year={2020},
  publisher={Elsevier}
}

@article{vinod2025projection,
  title={Projection-free computation of robust controllable sets with constrained zonotopes},
  author={Vinod, Abraham P and Weiss, Avishai and Di Cairano, Stefano},
  journal={Automatica},
  volume={175},
  pages={112211},
  year={2025},
  publisher={Elsevier}
}

@article{raghuraman2022set,
  title={Set operations and order reductions for constrained zonotopes},
  author={Raghuraman, Vignesh and Koeln, Justin P},
  journal={Automatica},
  volume={139},
  pages={110204},
  year={2022},
  publisher={Elsevier}
}

@article{koeln2020vertical,
  title={Vertical hierarchical {MPC} for constrained linear systems},
  author={Koeln, Justin and Raghuraman, Vignesh and Hencey, Brandon},
  journal={Automatica},
  volume={113},
  pages={108817},
  year={2020},
  publisher={Elsevier}
}

@article{andrade2024tube,
  title={Tube-based model predictive control based on constrained zonotopes},
  author={Andrade, Richard and Normey-Rico, Julio E and Raffo, Guilherme V},
  journal={IEEE Access},
  volume={12},
  pages={50100--50113},
  year={2024},
}

@inproceedings{siefert2023_SVSE,
  title={Set-valued state estimation for nonlinear systems using hybrid zonotopes},
  author={Siefert, Jacob A and Thompson, Andrew F and Glunt, Jonah J and Pangborn, Herschel C},
  booktitle={IEEE Conference on Decision and Control},
  pages={2172--2177},
  year={2023},
}

@inproceedings{thompson2025_MHE,
  title={Mixed-Integer Moving Horizon Estimation for Terrain-Aided Navigation Using Hybrid Zonotopes},
  author={Thompson, Andrew F and Robbins, Joshua A and Boler, Matthew E and Pangborn, Herschel C},
  booktitle={IEEE/ION Position, Location and Navigation Symposium},
  pages={815--820},
  year={2025},
  organization={IEEE}
}

@inproceedings{koeln2024zonolab,
  title={zonoLAB: A {MATLAB} toolbox for set-based control systems analysis using hybrid zonotopes},
  author={Koeln, Justin and Bird, Trevor J and Siefert, Jacob and Ruths, Justin and Pangborn, Herschel C and Jain, Neera},
  booktitle={2024 American Control Conference (ACC)},
  pages={2513--2520},
  year={2024},
  organization={IEEE}
}

@article{robbins2025sparsity_zonoopt,
  title={Sparsity-Promoting Reachability Analysis and Optimization of Constrained Zonotopes},
  author={Robbins, Joshua A and Siefert, Jacob A and Pangborn, Herschel C},
  journal={arXiv:2504.03885},
  year={2025}
}

@article{Braden01091986,
    author = {Bart Braden},
    title = {The Surveyor's Area Formula},
    journal = {The College Mathematics Journal},
    volume = {17},
    number = {4},
    pages = {326--337},
    year = {1986},
    publisher = {Taylor \& Francis},
}

@inproceedings{etsi_its_g5,
    author = {{ETSI}},
    publisher = {{European Telecommunications Standards Institute}},
    title = {Intelligent {Transport} {Systems} ({ITS}); {ITS-G5} {Access} layer specification for Intelligent Transport Systems operating in the {5~GHz} frequency band, {ETSI} EN 302 663 V1.3.1 (2020-01)},
    month = jan,
    year = {2020},
}

@inproceedings{etsi_cpm,
    author = {{ETSI}},
    publisher = {{European Telecommunications Standards Institute}},	
    title = {Intelligent {Transport} {Systems} ({ITS}); {Vehicular} Communications; Basic Set of Applications; {Analysis} of the {Collective} {Perception} {Service} ({CPS}), {ETSI} {TR} 103 562 V2.1.1 (2019-12)},
    month = dec,
    year = {2019},
}

@book{WHO2023global,
  title={Global status report on road safety 2023: summary},
  author={{World Health Organization}},
  year={2023},
  publisher={World Health Organization}
}

@inproceedings{Rezatofighi2019_IoU,
  title={Generalized intersection over union: A metric and a loss for bounding box regression},
  author={Rezatofighi, Hamid and Tsoi, Nathan and Gwak, JunYoung and Sadeghian, Amir and Reid, Ian and Savarese, Silvio},
  booktitle={Proceedings of the {IEEE/CVF} conference on computer vision and pattern recognition},
  pages={658--666},
  year={2019}
}

@article{polyhedra_vertex_enumeration_Assad2022,
author = {Assad, Caio and Morales, Gudelia and Arica, José},
year = {2022},
month = {08},
pages = {},
title = {Vertex Enumeration of Polyhedra},
volume = {42},
journal = {Pesquisa Operacional},
}

@inproceedings{Kopetzki2017,
	author = {Kopetzki, Anna-Kathrin and Schurmann, Bastian and Althoff, Matthias},
	title = {Methods for order reduction of zonotopes},
	year = {2017},
	booktitle = {2017 IEEE 56th Annual Conference on Decision and Control, CDC 2017},
	volume = {2018-January},
	pages = {5626 – 5633},
}

@misc{gurobi,
  author = {{Gurobi Optimization, LLC}},
  title = {{Gurobi Optimizer Reference Manual}},
  year = 2026,
  url = "https://www.gurobi.com"
}

@article{robbins2026_hybrid,
      title={Hybrid System Planning using a Mixed-Integer {ADMM} Heuristic and Hybrid Zonotopes}, 
      author={Joshua A. Robbins and Andrew F. Thompson and Jonah J. Glunt and Herschel C. Pangborn},
      year={2026},
      journal={arXiv:2602.17574},
}

\end{document}